\documentclass[%
 aip,
 pof,
 sd,%
 amsmath,amssymb,
preprint,%
]{revtex4-1}

\usepackage{graphicx}
\usepackage{amsfonts}
\usepackage{amssymb}
\usepackage{float}
\usepackage{latexsym}
\usepackage{amsmath}
\usepackage{subfigure}
\usepackage[euler]{textgreek}
\usepackage{textcomp}

\usepackage[table]{xcolor}

\newcommand{\tcbl}[1]{\textcolor{black}{#1}}

\newcommand{\eqr}[1]{(\ref{eq:#1})}

\newcommand{\eg}{\textit{e.g.}}
\newcommand{\bs}[1]{\boldsymbol{#1}}

\newcommand{\ie}{\textit{i.e.}}

\begin{document}

\title{\tcbl{Short-term oscillation and falling dynamics for a water drop dripping in quiescent air}}

\author{B.~Zhang}
\affiliation{Department of Mechanical Engineering, Baylor University, TX 76798, USA}

\author{P.-H.~Tsai}
\author{A.-B.~Wang}\email{abwang@iam.ntu.edu.tw}
\affiliation{Institute of Applied Mechanics, National Taiwan University, Taipei, 10617, Taiwan, R.O.C.}

\author{S.~Popinet}
\author{S.~Zaleski}
\affiliation{Sorbonne Universit\'es and CNRS, Institut Jean le Rond d'Alembert, UMR 7190, F-75005, Paris, France}

\author{Y.~Ling}\email{stanley\_ling@baylor.edu}
\affiliation{Department of Mechanical Engineering, Baylor University, TX 76798, USA}

%  P.-H.~Tsai\aff{2},
%  A.-B.~Wang\aff{2}\corresp{\email{abwang@iam.ntu.edu.tw}},
%  S.~Popinet\aff{3},
%  S.~Zaleski\aff{3},
% \and Y.~Ling\aff{1}
%  \corresp{\email{stanley\_ling@baylor.edu}}}

%\aff{2}Institute of Applied Mechanics, National Taiwan University, Taipei, 10617, Taiwan, R.O.C.
%\aff{3}Sorbonne Universit\'es and CNRS, Institut Jean le Rond d'Alembert, UMR 7190, F-75005, Paris, France}

\begin{abstract}
\tcbl{The short-term transient falling dynamics of a dripping water drop in quiescent air has been investigated through both simulation and experiment. A representative case with a low inflow rate in the dripping regime is considered. The focus is on the short term behavior and the time range considered covers about eight dominant second-mode oscillations of the drop after it is formed. Due to the small fluid inertia at the inlet, the growth of the drop is quasi-static and is well captured by the static pendant drop theory. Nevertheless, it is demonstrated that the pinching dynamics and the resulting post-formation state of the drop trigger a nonlinear oscillation when the drop falls. The initial shape of the drop when it is just formed is decomposed into spherical harmonic modes. The initial mode amplitudes, characterized by the Fourier-Legendre coefficients, are found to be finite for up to the tenth mode. The pinching dynamics such as interface overturning introduces small-scale variation on the drop contour, which in turn contributes to the finite amplitudes of the higher-order modes. Furthermore, the initial kinetic energy when the droplet is just formed is as important as the initial surface energy contained in the drop shape, and is found to amplify the initial oscillation amplitude and to induce a phase shift in the oscillation of all the modes.} By incorporating both the initial surface and kinetic energy, the linear model for a free drop oscillation yields very good predictions for the second and third modes. The mode amplitude spectra show both the primary frequencies that are consistent with the Lamb's theory and the secondary frequencies arising from different modes due to nonlinear inter-mode coupling. Moreover, it is worth to note that the nonlinear effect is most profound for the fourth mode owing to its resonant coupling with the dominant second mode. The complex transient flow inside and outside the drop is induced by the interaction between the falling motion and the nonlinear oscillation. The streamlines indicate that the internal flow is substantially different from the Hill vortex for a falling drop without oscillation. The temporal evolutions of both the internal flow and the wake morphology follow the dominant second oscillation mode. In the oblate-to-prolate deformation, the internal flow goes against the external flow. As a result, a saddle point arises in the drop, which gives rise to two counterrotating vortices. The vortex dynamics are visualized by the swirling-strength vortex identification criterion and the vorticity. Whereas the potential flow changes direction during a second-mode oscillation cycle, the rotating directions of the vortices remain the same.
\end{abstract}

\maketitle

\section{Introduction}
%problem description
%Dripping drops abound in nature and daily life, ranging from dripping faucet to inkjet printing \citep{Basaran_2013a, Delrot_2016a}. 
%\tcbl{More recently, the formation of the drop and the oscillation after the drop is formed have also been used to measure dynamic surface tension and viscosity \citep{MacLeod_1993a,Zhang_1994b,Staat_2017a}.}
%When liquid such as water is fed in a nozzle, a drop will develop at the nozzle exit and will eventually detach due to inertia or gravity. Depending on the flow speed in the nozzle, the drop formation can be divided into three different regimes: dripping, transition (or chaotic dripping), and jetting \citep{Clanet_1999a, Ambravaneswaran_2004a}. While the breakup in the jetting regime is mainly driven by the inertia of the injected liquid, for the dripping regime the acceleration of gravity plays the dominant role. In the present study, a specific case in the dripping regime is considered. 

The falling dynamics of an oscillating drop is essential to many natural phenomena and industrial applications, such as rain drops \cite{Feng_1991a} and inkjet printing \cite{Basaran_2013a}. For drops that are formed by a nozzle, the drop characteristics can be controlled through the inflow rate. When the inflow rate is large, the injected liquid inertia dominates and the drop formation is in the jetting regime; when the inflow rate is small, then the gravity plays the dominant role, placing the drop formation in the dripping regime \cite{Clanet_1999a}. In the present study, we focus on one specific case in the dripping regime. The purpose of the study here is to provide a comprehensive description of the short-term oscillation and falling dynamics for the dripping drop. 

The initial conditions for the drop fall are determined by the drop formation process. Since the shape oscillation of the falling drop is triggered by the non-equilibrium shape and the velocity field when the drop is just formed. The shape oscillation will in turn impact the falling dynamics of the drop and the development of the transient flow around the drop. 
Nevertheless, despite its importance, the effect of drop formation on the subsequent oscillation and falling dynamics have not received enough attention in former studies. 
Instead of using the precise post-drop-formation state, \textit{ad hoc} initial conditions 
(such as simple spheroid shape), are often used in simulations \citep{Lalanne_2013a,Agrawal_2017a,Bergeles_2018a}. 
To fully incorporate the effect of drop formation, the whole process starting from drop growth, 
continuing with detachment, and eventually fall, is considered in the present simulation. 
Another important advantage of simulating the whole process is that an experiment with 
exact conditions can be done to validate the simulation results. This is hard to achieve if \textit{ad hoc} initial conditions are specified like in former simulations. 

\subsection{Drop formation}
The dripping drop first develops as a pendant drop, hanging at the nozzle exit. When the drop volume is smaller than the critical volume, the surface tension is strong enough to resist gravity and to keep the drop stably attached to the nozzle \citep{Padday_1973a, Sumesh_2010a}. As the volume of the pendant drop reaches the critical value, 
the drop becomes unstable and a neck is formed between the nozzle and the main body of the drop \citep{Schulkes_1994a,Coullet_2005a}. 
The minimum radius of the neck rapidly decreases, giving rise to an increasingly large capillary pressure in the neck. This high pressure drives the liquid away from the neck toward the nozzle and the main body of the drop, further accelerating the pinching process. 
%The neck then turns into a thin liquid bridge. 

The pinching of the liquid neck will eventually detach the drop \tcbl{from the nozzle} and the pinching dynamics has been studied 
extensively in the past. The overall pinching process is dictated by surface tension, inertial, and viscous forces \citep{Castrejon-Pita_2015a}.
 The pinching process exhibits a finite-time singularity and a universal self-similar behavior near the singularity \tcbl{\citep{Eggers_1993a, Eggers_1994a, Papageorgiou_1995a, Day_1998a, Zeff_2000a, Chen_2002a, Doshi_2003a, Castrejon-Pita_2012a}}.
For low-viscosity liquids like water, 
inertia of the liquid flow toward \tcbl{the} main body of the drop results in the shift of the local minimum of bridge radius toward the top of the drop, where the interface overturns before pinching eventually occurs.

%Though the Marangoni effect is not considered in the present study, it has been reported that the capillary pressure gradient can destabilize a pendant drop in the absence of gravity  \citep{Suryo_2006a}. 
To obtain details of the flow field in the drop formation process, advanced experimental diagnostics and high-resolution simulations are required \citep{Wilkes_1999a, Bos_2014a, Borthakur_2017a}. 
By recording two consecutive images of the same drop with a small time delay, \citet{Bos_2014a} extracted the \tcbl{longitudinal} velocity profile during drop formation. 
%The contraction of filament has also been shown to have an impact on drop formation at the tip \citep{Schulkes_1996a, Notz_2004a,Hoepffner_2013a}. For example, the experimental and numerical studies by \citet{Hoepffner_2013a} showed that the creation of a vortex ring by a Venturi jet through the neck, by which the neck can escape from pinch-off. 
\tcbl{ 
%Accurate measurement of the drop formation process can lead to estimate the dynamic surface tension of liquids \citep{MacLeod_1993a,Zhang_1994b}.
For the present problem, the viscosity and density of the surrounding air are small compared to those for water, and the effect of the surrounding air on drop formation is small. When the surrounding fluid has similar density or viscosity as the drop fluid, the surrounding fluid can  have a significant impact on the drop formation dynamics \citep{Zhang_1999b}. } 

\subsection{\tcbl{Oscillation of a free drop}}
%former work on free oscillation
Following the formation, the drop falls under the action of gravity. Since the shape of the drop just after detachment is out of equilibrium, the capillary force will cause the drop to oscillate when it falls. 
Drop oscillation is a classic fluid mechanics problem, and the early investigation on the oscillation of a free drop can be traced back to the pioneering work 
of \citet{Rayleigh_1879a}. 
\tcbl{(A free drop here is referred to a drop that is located in an unbounded domain without gravity and falling motion.)} 
For the infinitesimal amplitude oscillation of a free and inviscid liquid drop, Rayleigh decomposed the shape of the drop into spherical harmonic modes and calculated the corresponding frequency for each mode \citep{Rayleigh_1879a}. The original work of Rayleigh is based on a free-surface approximation. 
The extension to incorporate the effect of ambient fluid and the viscous effect was made by \citet{Lamb_1932a}, and later followed by others \citep{Reid_1960a, Miller_1968a,Prosperetti_1980a}. 
\tcbl{
Lamb's theory is generally valid for low-viscosity fluids. 
Yet \citet{Miller_1968a} showed that even if the viscosities of the drop and surrounding fluid are both small, 
the viscous effect cannot be ignored since the oscillation damping rate is controlled by the boundary layer developed near the interface. The transient effect on the oscillation frequency and the damping rate was investigated by \citet{Prosperetti_1980a} and it is shown that the predictions based on normal mode analysis by \citet{Lamb_1932a} are strictly valid only asymptotically.
When the oscillation amplitude is finite, the nonlinear effect on drop oscillation becomes important. Typical nonlinear effects include  decrease of oscillation frequency with oscillation amplitude, asymmetry in oscillation amplitude, and coupling between different oscillation modes \citep{Tsamopoulos_1983a, Natarajan_1987a, Becker_1991a, Basaran_1992a, Becker_1994a}. 
}

\subsection{Dynamics of a falling drop}
%former work on oscillation of a falling drop
For a falling drop, the oscillation dynamics and the \tcbl{transient} flow around the drop
become more complicated. 
Extensive numerical and experimental studies have been performed to understand the long-term falling dynamics 
of liquid drops after the terminal velocity is reached  \tcbl{(see for example \citet{Gunn_1978a, Feng_1991a, Helenbrook_2002a, Feng_2010a})}. Those research efforts were usually 
motivated by the interest in rain drops in atmospheric science. 
The present study has a different focus, that is, on the short-term dynamics of the falling drop. \tcbl{Here, the short-term and long-term are defined 
with respect to the response time required for the drop to reach the terminal velocity.} 
The interest on the short-term behavior is motivated by the fact that, for many application of falling drops, such as inkjet printing, the drop will reach a substrate or a liquid film far before reaching the quasi-steady state. Furthermore, the oscillation dynamics of a falling drop in the short term has also lead to new technology to measure liquid properties, \eg, \citet{Staat_2017a} recently proposed new methods to measure surface tension and drop viscosity based on the short-term oscillation frequency and damping rate.

In the short term, the drop velocity and Reynolds number increase over time and the viscous flow around the drop is transient. 
Furthermore, due to the falling motion and the induced shear stress, the equilibrium shape of the oscillating drop 
is not spherical in general \citep{Feng_2010a}. 
Because of these additional complexities, there is no general analytical solution for the problem 
and numerical approaches are required to solve the governing equations
\citep{Lalanne_2013a, Tripathi_2014a, Agrawal_2017a, Bergeles_2018a}. 
Owing to the similar dynamics between a falling drop and a rising bubble, these two cases are often discussed together (see for example \citet{Ern_2012a}), although fundamental difference between these two cases exists \citep{Tripathi_2014a}.
It is challenging to accurately measure the three-dimensional flow inside a small drop in experiments. 
By seeding tracer particles of an average size of 10 \textmu m, \citet{Chung_2000a} obtained instantaneous velocity maps inside an oscillating drop which is electrostatically levitated. 

\subsection{Numerical Simulation}
Thanks to the development of advanced interface capturing techniques in the past decades, direct numerical simulation is now capable of capturing interfacial flows that exhibit topology changes \citep{Tryggvason_2011a} and can also provide high-level details of the flow field that are difficult to measure in experiments. 
Extensive numerical studies have been conducted to simulate the drop formation process by the volume-of-fluid (VOF) method, see for example \citet{Zhang_1999a} and \citet{Gueyffier_1999a}. 
The recent simulations by \citet{Agrawal_2017a} have used the VOF method to resolve the oscillation of a falling drop with a non-spherical initial shape. 
It is shown that the oscillation only arises in the longitudinal direction and no azimuthal variation 
was observed even when vortex shedding occurs in the wake of the drop. 
Another recent work by \citet{Bergeles_2018a} presented high-resolution three-dimensional simulation results 
for a falling drop of millimeter class. The detailed flow structure was well captured and in particular, the roller vortex that is required to link the circulation in the wake of the drop with a Hill vortex inside the drop was clearly unveiled. 
For a similar problem, \citet{Lalanne_2013a} have performed axisymmetric simulations using the level-set method
for  the oscillation of rising drops and bubbles. It was found that the oscillation frequency decreases slightly 
with the rising velocity while the damping rate of the drop oscillation is significantly magnified due to the rising motion. 

\subsection{Goal of this study}
In spite of of the extensive studies discussed above, a comprehensive understanding of the 
short-term oscillation and falling dynamics for a dripping drop remains to be established. 
In particular, the effect of drop formation on the oscillation dynamics and the transient flow around the falling drop are still not fully understood. 
To the knowledge of the authors, there is no previous study that considers 
the effect of the drop formation on the oscillation dynamics of a falling drop. 
The oscillation of a drop is dictated by the initial conditions 
which are in turn set by the drop formation process. 
Former numerical studies generally assumed the initial drop shape to be ellipsoidal or spherical 
with a constant initial velocity within the drop \citep{Lalanne_2013a,Agrawal_2017a,Bergeles_2018a}. 
However, the shape of the drop when it is just formed is far more complex than an ellipsoid, 
\tcbl{and furthermore, the velocity field in the just-formed drop is highly non-uniform due to the pinching dynamics.}  
The former simulations with simplified initial conditions are useful to understand the general physics 
of oscillation of a falling drop. Nevertheless, in order to precisely predict the shape and dynamics of the falling drop, which are critical to many applications of drops, 
\eg, the impact of a falling drop on a deep pool \citep{Deka_2017a},  
the effect of drop formation on the subsequent drop oscillation and falling dynamics must be faithfully incorporated. 

The goal of the present study is therefore to investigate the dynamics of a water drop dripping in quiescent air through simulation and experiment.
Particular focus will be placed on the drop oscillation dynamics and the development of the transient flow around the drop. 
To achieve this goal, one specific case is considered in the present study. 
The flow rate at the nozzle inlet is chosen to be sufficiently small, 
so that the drop formation is in the dripping regime and the drop growth is quasi-static.
Furthermore, we focus on only the short term of the drop fall, 
during which the drop shape and the flow remain axisymmetric. 
%During the time range considered, the drop velocity is far from the terminal velocity. As a result, the flow is transient. 
%We note however that the coupled experimental and numerical investigations performed here take into account viscous effects 
%and could be used to revisit correlations on drop volume developed in former studies \citep{Scheele_1968a, Zhang_1995a}.
The key questions that the present study aims to address include:
\begin{itemize}
	\item \tcbl{Are the ``initial conditions" set by the drop formation process important to the drop oscillation and falling dynamics? }
	\item \tcbl{How do the nonlinear dynamics and falling motion influence the drop oscillation dynamics, such as the oscillation frequency and damping rate?}
	\item How do the drop oscillation and the falling motion contribute to the development of the transient flow around the drop? Is the flow structure within the drop similar to the classic Hill vortex? 
\end{itemize}
To address these questions, axisymmetric simulations are carried out with the adaptive multiphase flow solver, \emph{Gerris}. An experiment with the same conditions has also been conducted to validate the simulation results. The simulation and experimental approaches are described in section \ref{sec:methods}. The results for drop formation, shape oscillation, and transient flow around the drop, will be presented and discussed in sequence in sections \ref{sec:results_formation}, \ref{sec:results_oscillation} and \ref{sec:results_transient}, respectively. Finally, concluding remarks will be given in section \ref{sec:conclusions}. 

\section{Methodology}
\label{sec:methods}
\subsection{Key parameters}
The process of drop formation are controlled by physical parameters listed in table \ref{tab:parameter}, including the liquid and gas properties, the nozzle radius, the gravity acceleration, and the inlet flow rate. The mean inflow velocity, $u_0=Q/\pi R_0^2=0.265$ mm/s can serve 
as an alternative for the inflow rate $Q$.  %u_0=0.265$ mm/s
The key dimensionless parameters can be derived and the values are given in table \ref{tab:dimension1}. Since the gas-to-liquid density and viscosity ratios, $r$ and $m$, are both very small, the effect of the gas phase on drop formation is small. 
The Weber, Ohnesorge, and Bond numbers \tcbl{are} measures of the relative importance of the fluid inertia, liquid viscosity, and gravity with respect to surface tension.
For the  small $Q$ used in the present problem, the drop formation process is quasi-static and $We=8.17\times10^{-7}\ll 1$. 
The effect of inflow inertia is thus negligible. The variation of $We$ does not influence the drop formation  \citep{Wilkes_1999a}
and the value of $Q$ is immaterial to the results to be presented, as along as $Q$ remains to be small. 
Due to the relatively low viscosity of water, $Oh=0.00426$, is also very small, 
suggesting that the viscous effect is generally small in the drop formation process. 
Finally, the Bond number is the primary dimensionless parameter 
to determine the sizes of the detached primary and secondary drop.

\begin{table}
\centering
\begin{tabular}{ccccccccc} 
     \hline
       $\rho_l$    &$\rho_g$    &$\mu_l$       &$\mu_g$      &$\sigma$      &$R_0$        &$g$           &$Q$                         \\
        (kg/m$^3$)     &(kg/m$^3$)      &(Pa s)           &(Pa s)     &(N m)        &(m)        &(m/s$^2$)        &($\mu$L/min)   \\[0.5em]
    \hline
                 1000               &1.2                &$0.85 \times 10^{-3}$         &$1.8 \times 10^{-5}$       &0.0688        &$8 \times 10^{-4}$                  &9.81        &32                    \\
    \hline
\end{tabular}
\caption{Physical parameters for the formation of a dripping drop.}
\label{tab:parameter}
\end{table}

%\tcbl{
%Other dimensionless parameters can be derived based on those listed in Eq.\ \eqr{dmls_para_list}. 
%The Reynolds number for the inflow, $Re=\sqrt{We}/Oh=0.21$, the small value 
%here indicates the fluid inertia is small compared to viscosity. 
%The Froude number $Fr=\sqrt{We/Bo}=0.00299$, which is very small, suggesting that the gravity 
%is the dominant destabilization force to the pendant drop. 
%The capillary number is a different way to measure the ratio between viscous stress 
%and surface tension and is also commonly used to characterize drop formation \citep{Utada_2007a}. 
%Due to the small $Fr$, the gravitational velocity, $(gR_0)^{1/2}$, instead of $u_0$, serves as the characteristic velocity scale. Therefore, the capillary number can be defined as $Ca \equiv {\mu_l\sqrt{gR_0}}/{\sigma}$ and it can be shown that $Ca= Oh \sqrt{Bo}$. 
%The small value, $Ca=0.00128$, simply affirms that the viscous effect is generally negligible  
%in the drop formation. 
%}

\begin{table}
\centering
\begin{tabular}{ccccc} 
    \hline
        $r$	& 	$m$	& $We$          &	$Oh $    & $Bo$     \\
        $\rho_g/\rho_l$ & $\mu_g/\mu_l$  & $\rho_l u_0^2 R_0/\sigma$   &$\mu_l/\sqrt{\rho_l\sigma R_0}$ & $\rho_lgR_0^2/\sigma$   \\
    \hline
             0.0012	&   0.021	 &  $8.17\times 10^{-7}$ & 0.00426                 & 0.091              \\
    \hline
\end{tabular}
\caption{\tcbl{Key dimensionless parameters for the drop formation. }}
\label{tab:dimension1}
\end{table}

After the drop detaches from the nozzle, the drop radius is measured to be $R_d=1.86$ mm. 
The oscillation and falling dynamics of the drop can be fully determined 
by the Reynolds and Weber numbers based on the drop diameter ($D_d=2R_d$), 
namely $Re_d\equiv D_d u_d\rho_g/\mu_g$ and $We_d\equiv D_d u_d ^2\rho_g/\sigma$,
along with the post-formation state of the drop as the initial conditions. 
As the drop velocity, $u_d$, increases over time, $Re_d$ and $We_d$ rise
accordingly. In the time range considered in the present study,
the drop velocity increases from 0.07 m/s (just after detachment) to about 1.70 m/s. 
The corresponding range of drop Reynolds and Weber numbers are $25.9 \lesssim Re_d \lesssim 633$ 
and $2.62 \times 10^{-4} \lesssim We_d \lesssim 0.156$. For this 
range of $Re_d$, the flow was observed to remain approximately axisymmetric in the experiment. 
It was measured that the deviation  
of the drop centroid from the nozzle axis is smaller than 0.3\% of the falling distance in the time range considered. 
Furthermore, as $We_d$ is small, the surface tension will be sufficient to avoid an aerobreakup. 
According to the experiment of \citet{Gunn_1949a}, the terminal velocity for this drop size is about 
6.2 m/s. The Reynolds and Weber numbers corresponding to the terminal falling velocity will then be about 
$Re_{d,t=\infty}\approx1600$ and $We_{d,t=\infty}\approx2.4$. It is clear that the drop velocity in the present study remains far from the terminal state. 
The drop oscillation Ohnesorge number, $Oh_{osc}=\mu_l/(\rho_l\sigma R_d)^{1/2}$, 
is often used to characterize the viscous effect on the oscillation of a free drop, 
which can be expressed as $Oh_{osc} = \sqrt{2We_d} /(m Re_d)$. 
(Alternatively, the oscillation Reynolds number, $Re_{osc}=1/Oh_{osc}$, can be used.)
Here $Oh_{osc}=0.00278$, is very small, therefore, it is expected the viscous effect on the drop oscillation is small. 

\tcbl{Due to the rich flow physics involved in the present problem, we have focused on only one specific case, 
instead of a parametric study. If the key dimensionless parameters listed in Table \ref{tab:dimension1} vary, 
the specific values in the results to be shown later will change. However, the case selected here well represents 
millimeter-size low-viscosity droplets in the dripping regime. The conclusions with regard to 
the droplet formation, oscillation, and falling dynamics will remain valid  
as long as both of the Ohnesorge and Bond numbers are significantly smaller than unity. 
Parametric study for wider ranges of $Oh$ and $Bo$ is of interest but will be relegated to future work.}

 \subsection{Modeling and Simulation}
 
 \subsubsection{Governing equations}
The one-fluid approach is employed to resolve the two-phase flow, 
where the phases corresponding to the water drop and the ambient air
are treated as one fluid with material properties that change abruptly across the interface. 
The incompressible Navier-Stokes equations 
with surface tension can be written as
\begin{align}
	\rho (\partial_t\bs{u} + \bs{u} \cdot \nabla \bs{u}) = -\nabla p 
	+ \nabla \cdot (2\mu \bs{D}) + \sigma \kappa \delta_s \bs{n}\, , 
	\label{eq:mom} \\
	\nabla \cdot \bs{u} = 0 \, ,
	\label{eq:divfree}
\end{align}
where $\rho$, $\mu$, $\bs{u}$, and $p$ represent density and viscosity, velocity and pressure, respectively. 
The strain-rate tensor is denoted by $\bs{D}$ with components $D_{ij}=(\partial_i u_j + \partial_j u_i)/2$. 
The third term on the right hand side of Eq.\ \eqr{mom} is a singular term, with 
a Dirac distribution function $\delta_s$ localized on the interface, and it represents 
the surface tension. The surface tension coefficient is $\sigma$, 
and $\kappa$ and $\bs{n}$ are the local curvature and unit normal of the interface. 

The liquid volume fraction $C$ is introduced to distinguish the different phases, 
in particular $C=0$ in the computational cells with only air (respectively $C=1$ in cells containing only water), and its time evolution satisfies the advection equation
\begin{align}
	\partial_t C + \bs{u} \cdot \nabla C =0 \, .
	\label{eq:color_func}
\end{align}
The fluid density and viscosity are then determined by
\begin{align}
	\rho  & = C \rho_l + (1-C) \rho_g \, , 
	\label{eq:density} \\
	\mu  & = C \mu_l + (1-C) \mu_g \, ,
	\label{eq:viscosity}
\end{align}
where the subscripts $g$ and $l$ represent the gas phase (air) and the liquid phase (water), respectively. 

\subsubsection{Numerical methods}
\label{sec:numerics}
The Navier-Stokes equations (Eqs.\ \eqr{mom} and \eqr{divfree}) are solved by
the open-source solver \emph{Gerris} \citep{Popinet_2003a,Popinet_2009a}. 
In \emph{Gerris}, a finite-volume approach based on a projection method is employed.   
A staggered-in-time discretization of the volume-fraction/density and pressure 
leads to a formally second-order accurate time discretization. 
The interface between the different fluids are tracked 
by solving the advection equation (Eq.\ \eqr{color_func}) 
using a Volume-of-Fluid (VOF) method \citep{Scardovelli_1999a}. 
A quadtree spatial discretization is used, which gives a very important flexibility 
allowing dynamic grid refinement into user-defined regions. 
Finally the height-function (HF) method is used to calculate the local interface curvature, 
and a balanced-force surface tension discretization is used \citep{Francois_2006a, Popinet_2009a}. 
 
\subsubsection{Simulation setup}

 \begin{figure}
\begin{center}
\includegraphics [trim=0 0.5in 0 0.5in,clip, width=1.0\columnwidth]{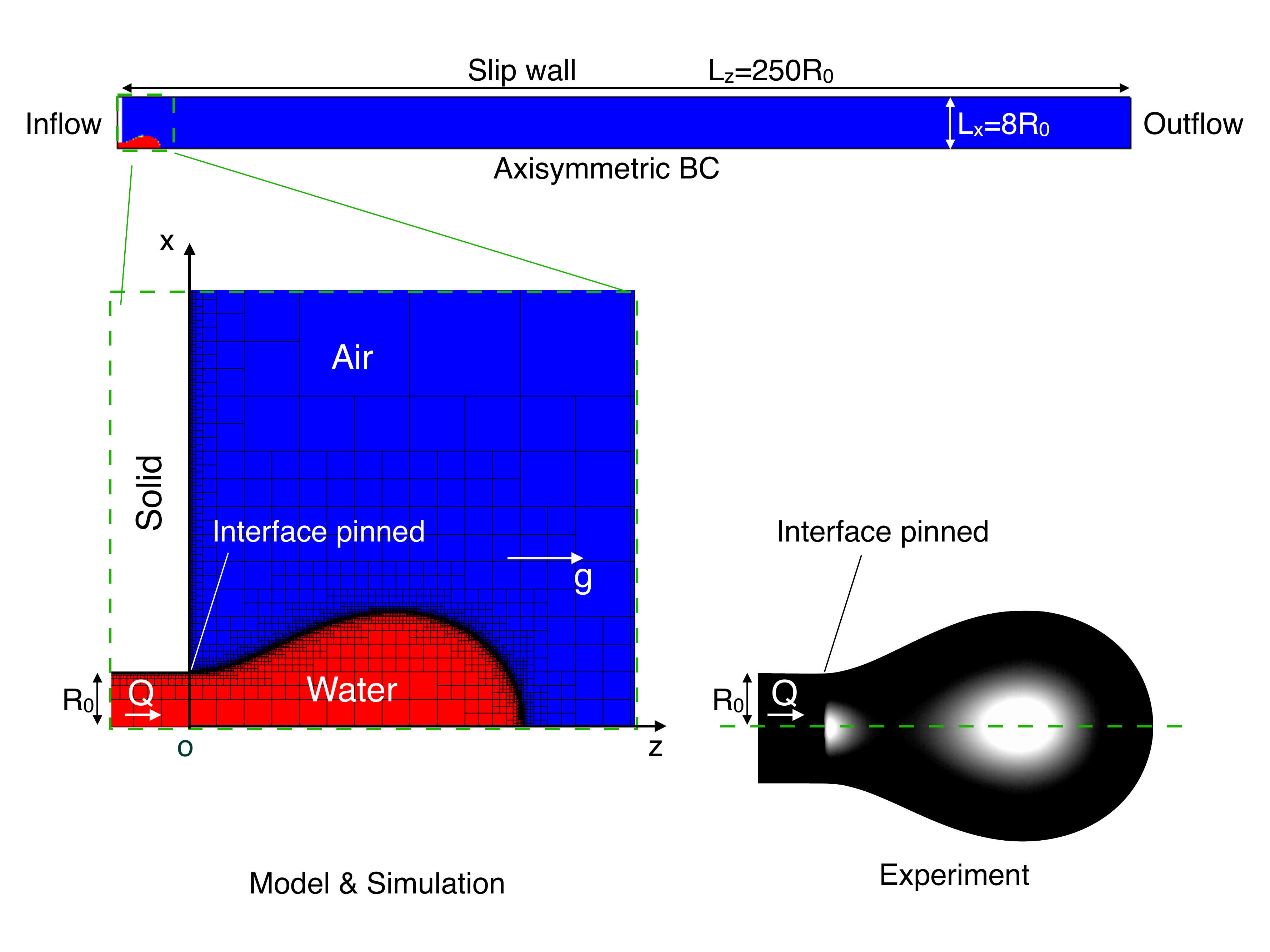}
\caption{Simulation setup.  }
\label{fig:simulation}
\end{center}
\end{figure}

In the numerical model, the flow is assumed to be axisymmetric. The 2D computational domain is shown 
in  figure \ref{fig:simulation}. The gravitational acceleration is along the $z$ direction. The water is injected into the domain 
from the left and the inlet flow rate $Q$ is kept the same as in the experiment. 
The thickness of the nozzle wall is ignored in the model. 
The ratio between the inner and outer radii of the nozzle in the experiment is 0.75. It has been shown by \citet{Ambravaneswaran_2002a} that the nozzle wall thickness can affect the drop formation dynamics when the flow rate is high. For the present problem, a very small flow rate has been used. According to the experimental results of \citet{Zhang_1995a} for similar small flow rates, the effect of the wall thickness becomes negligible if the ratio of the inner to the outer radii of the nozzle exceeds 0.2. The ratio in the present experiment is significantly larger than the critical value and thus the effect of nozzle wall thickness on the drop formation can be ignored. 

Furthermore, a solid block is added above the nozzle, 
see figure \ref{fig:simulation}. The boundary condition of the volume fraction $C$ at the solid boundary is $\partial C/\partial n=0$. \tcbl{The reason for adding the solid block is to pin the contact line, where water, air, and solid meet, at the block corner. In the experiment, the interface is pinned 
at the outer perimeter of the nozzle. By setting the distance between this pinned point to the $z$--axis as $R_0$, the model and the experiment 
exhibit the same $Bo$. In both experiment and simulation, the contact angle varies slightly when pinching occurs, 
and the contact line remains pinned during the drop formation process. }

Thanks to the adaptive mesh, a computational domain that is significantly larger than the drop size can be used.  
As a result, the effect of boundaries on the drop can be eliminated. The length of the domain is
 $L_z=200\ \mathrm{mm}=250 R_0$ and the height is $L_x=6.4\ \mathrm{mm}=8 R_0$. The axisymmetric boundary condition 
 is invoked at the bottom of the domain. The inflow \tcbl{(Dirichlet velocity and Neumann pressure conditions)} and outflow \tcbl{(Dirichlet pressure and Neumann velocity conditions}) BC's are applied to the left and the right of the domain. The top boundary is considered as a slip wall. The minimum cell size used in the simulation is determined by the maximum mesh refinement level, $L$, namely $\Delta_{\min}=L_x/2^L$. 
Different refinement levels have been tested and the grid-refinement results are to be shown in the next section. 
The time step is computed based on the restriction from the advection, viscous, and surface tension terms in the governing equations. For the present problem, the time step restriction is mainly from the surface tension due 
to the small capillary number \cite{Ling_2016b}.  
 
 \subsection{Experiment}
 \begin{figure}
\begin{center}
\includegraphics [width=8.6 cm]{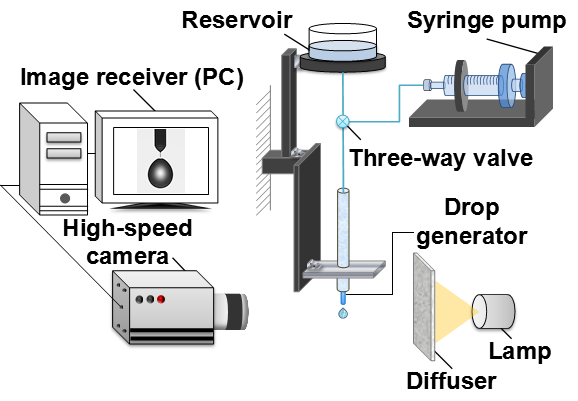}
\caption{Experimental setup.}
\label{fig:setup}
\end{center}
\end{figure}

Figure \ref{fig:setup} shows the experimental setup to investigate the formation and the fall of the water drop using high-speed imaging. A stainless steel nozzle with sharp-edged exit surface was used, and its inner and outer radii are 0.6 mm and 0.8 mm, respectively. Water drops were then generated from the nozzle either by the pressure from a constant-height reservoir, or by the pressure from a syringe pump (KDS210, KD Scientific). High-speed camera (NAC Memrecam GX-1) with frame rates varying from 100 to 5,000 fps (frame per second) have been used to capture the shape of the drop. The spatial resolution and exposure time varies in the range of 20-70 $\mu$m/pixel and 20-200 $\mu$s, respectively. To minimize the influence of vibrational disturbances and temperature variations in the pinching of the drop, all experiments were conducted on an anti-vibration table in the isolated corner of a basement with air-conditioning. For better visualization, uniform illumination was achieved by placing a diffuser in front of the 100W white light LED lamp. \tcbl{To avoid the heating effect, the LED light was placed 1.5 meter away from the observation area and the LED light was turned on only during recording.} The images obtained by high-speed camera were post-processed by Matlab code to measure the geometric properties of the drop before and after detachment, such as the volume and height of the pendant drop, the radius of the neck, the eccentricity of the falling drop. 

Surface tension was measured by Du No\"uy ring method. Temperature (25 $^{\circ}$C) and density of the test liquid were measured by a temperature recording device (Chino AH3760 with Pt100 sensor) and a mass-volume method, respectively. Liquid viscosity was determined using a rotational viscometer (Brookfield DV-II).

 \section{Results for drop formation}
 \label{sec:results_formation}
 
 The focus of the present study is on the oscillation and falling dynamics after the drop is detached 
 from the nozzle. Nevertheless, since we aim at unveiling the effect of drop formation on the 
 subsequent shape oscillation, the results for the drop formation will be first presented and validated
 against theory and experiment. 
 
 \subsection{General process and time scales}
\begin{figure}
\begin{center}
\includegraphics [width=1.0\columnwidth]{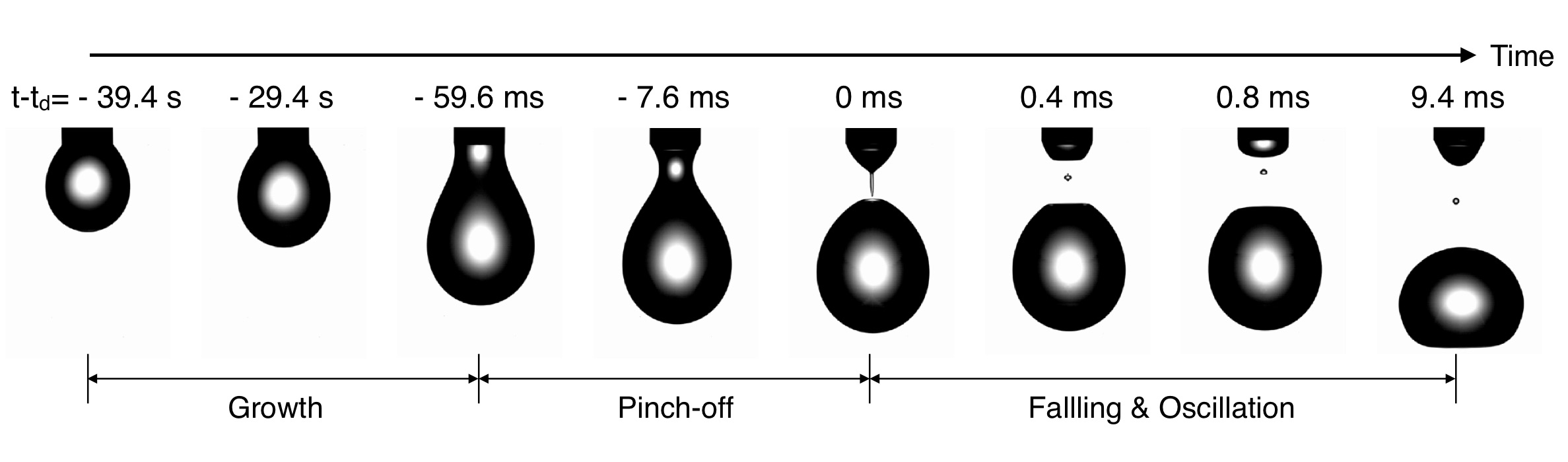}
\caption{Overall process of a drop dripping from a nozzle: growth, pinch-off, and fall shown by high-speed camera images.}
\label{fig:evolution}
\end{center}
\end{figure}

A sequence of images of the drop obtained from high-speed imaging are shown in figure \ref{fig:evolution} 
to depict the process of drop formation and subsequent fall in quiescent air. 
The overall process can be generally divided into three phases: 
growth, pinch-off, and fall. When the drop falls, it deforms in an oscillatory manner. 

It should be noted that the time scales for different phases in the process are different. 
(The time differences between the images shown in figure \ref{fig:evolution} are not even.) 
The growth of the drop is very slow compared to the other two phases, 
simply due to small flow rate at the nozzle inlet. 
It takes about one minute for the pendant drop to grow to the critical volume. 
In contrast, when the drop volume reaches the critical volume, 
the developing and pinching of the neck of the pendant drop evolve at a very fast speed,  
taking about a millisecond. 
When the detached drop falls in air, the dominant oscillation period is about $\tau_{osc}=21.5$ ms. 
This multiple time-scale nature makes the investigation challenging for both experiment and simulation 
if one aims at capturing the whole process from drop formation to fall. 

To overcome this challenge, \tcbl{multiple experiments with different frame rates were conducted} to capture 
different phases. For the growth of the drop, a low frame rate, 100 fps was used.  For the pinching and oscillation, a high frame rate, 5000 fps was used. The theoretical solution of a static pendant drop that is close to the critical volume is used to initialize the simulation. The initial velocity throughout the domain is taken to be zero since the pendant drop is quasi-static. For the most refined simulation ($L=11$), the simulation starts at the time that is 394 ms before the drop detaches, namely $t_d-t=394$ ms. 
%Validation of the simulations will be shown in the following section. 

 \subsection{Drop growth as a pendant drop}
 \label{sec:growth}
\begin{figure}
\begin{center}
\includegraphics [width=0.5\columnwidth]{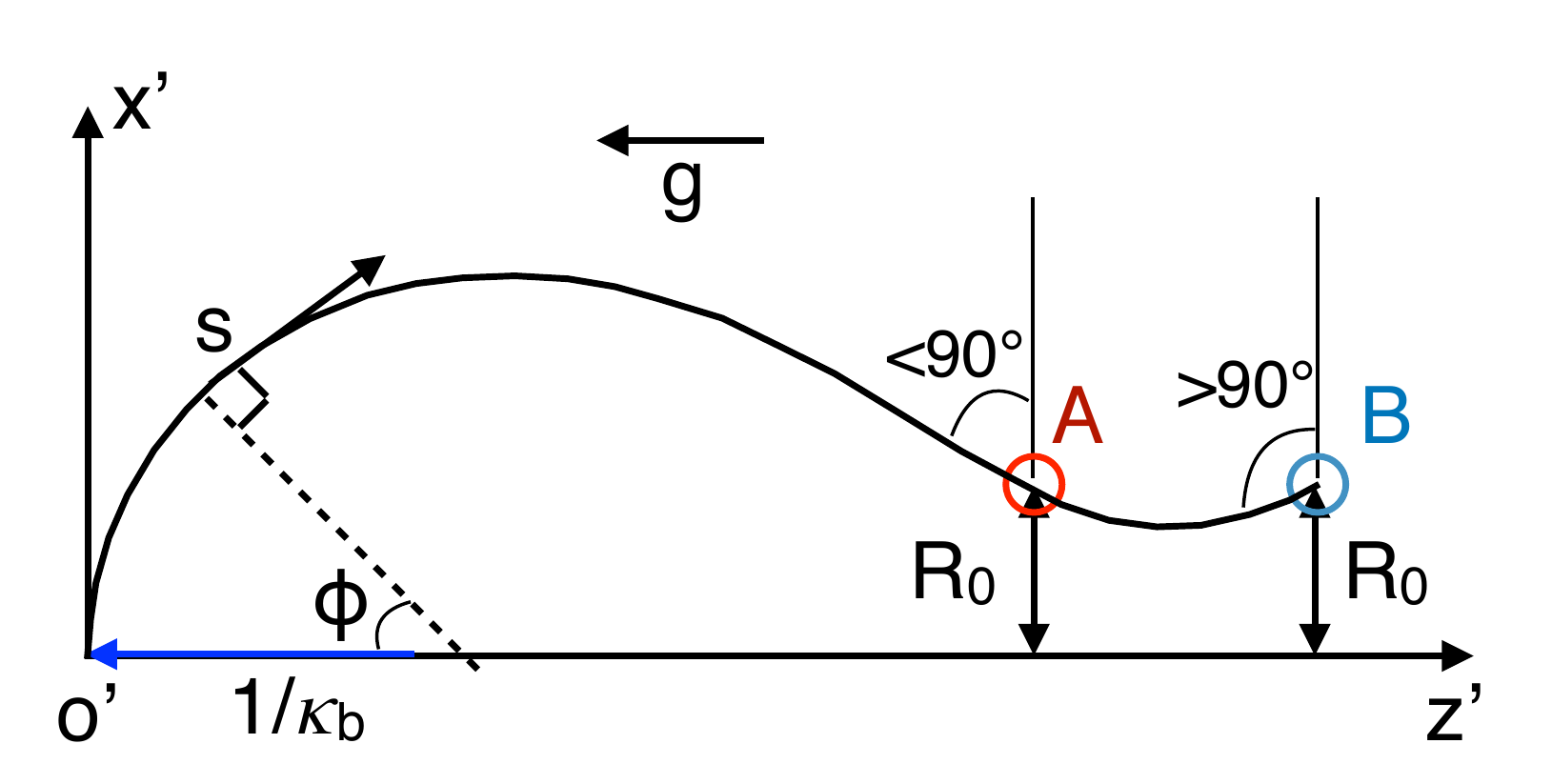}
\caption{Sketch of the axisymmetric quasi-static pendant drop profile.}
\label{fig:static}
\end{center}
\end{figure}

\tcbl{Due to the small Weber and Ohnesorge numbers} in the present problem, the effects of liquid inertia 
and viscosity on the drop formation are negligible compared to that of the surface tension.  
As a consequence, the drop grows quasi-statically and follows the static pendant drop theory \citep{Padday_1973a}.
For a static pendant drop, its shape is axisymmetric and the surface tension and the gravitational force 
are in  equilibrium. 
The shape of the drop can then be obtained by solving a set of ordinary differential equations, 
which are given in Appendix \ref{sec:pendant_drop_theory}. The integration of the equations 
is from the bottom of the pendant drop as shown in figure \ref{fig:static}, (a new coordinate $(x',z')$ is used,)
with the curvature at the drop bottom $\kappa_b$ as the boundary condition. 
For each $\kappa_b$, there are multiple solutions that satisfy a given Bond number \citep{Coullet_2005a}. 
Here only the two solutions which give drop volumes which are close to the critical volume are relevant. The two solutions are schematically shown in figure \ref{fig:static}. While for solution A the angle between the interface and the nozzle exit is less than 90\textdegree, for solution B the angle is larger than 90\textdegree. 

 \begin{figure}
\begin{center}
\includegraphics [width=1.0\columnwidth]{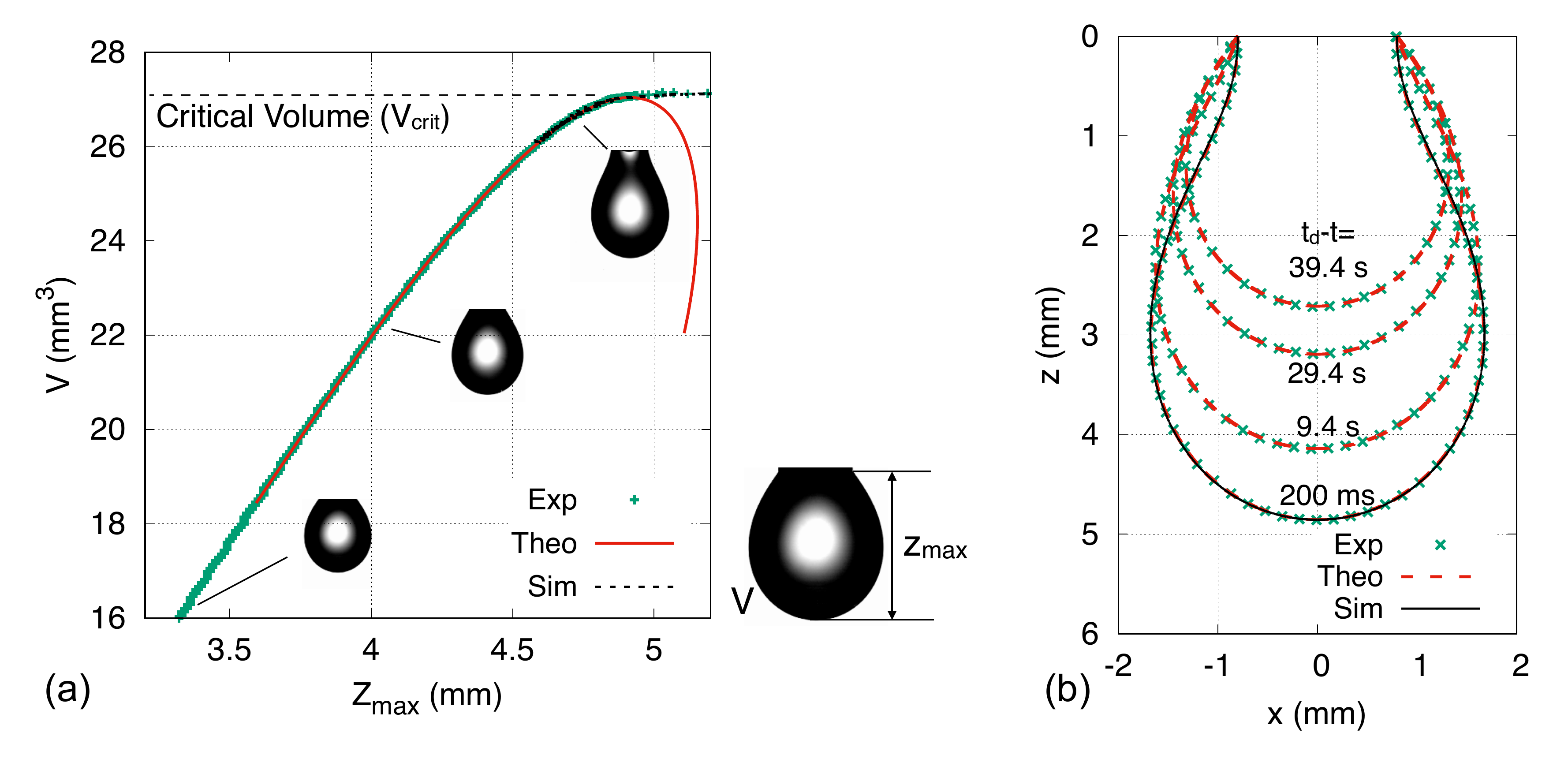}
\caption{Comparison of static pendant drop theory with experimental and simulation results: (\textit{a}) drop volume $V$ versus drop height $Z_{\max}$; (\textit{b}) drop contours at different times. \tcbl{The critical volume shown in (a) is $V_{crit} = 27.05$ mm$^3$.} } 
\label{fig:volume}
\end{center}
\end{figure}

The volume ($V$) and the height ($Z_{\max}$) of the pendant drop can be measured from the experimental and numerical results, which are shown along with the pendant-drop theoretical predictions in figure \ref{fig:volume}(a). It can be observed the experimental and theoretical results agree very well before the drop volume reaches the critical volume. The critical volumes measured from the experiment and simulation are both about 27.10 mm$^3$, which is very close to the value predicted by the pendant-drop theory, \ie, $V_{crit}=27.05$ mm$^3$.  The $V$-$Z_{max}$ curves obtained in the experiment and simulation appear to be flat when pinching occurs. During the pinching process, the rapid increase of $Z_{max}$ is due to the redistribution of volume within the drop, as a result, the drop volume increase in the fast pinching process is negligibly small. 
The initial conditions for the simulation are taken from the theoretical result for $V=26$ mm$^3$. At the time, 
the angle between the interface and the nozzle exit is less than 90\textdegree\ (case A in figure \ref{fig:static}). If the inflow at the nozzle is stopped, the pendant drop will remain stable. 
The simulation results of the $V$-$Z_{\max}$ curve at later times match very well with both the experiment and theory, see figure \ref{fig:volume}(a). This validates the present simulation setup in capturing the drop growth. \tcbl{The experimental and numerical results deviate from the theoretical solution beyond the critical $z_{\max}$, since the latter represents unstable static solution which will not be observed in reality. }

The excellent agreement between the experimental and theoretical results are also achieved in the contours of the drop 
at different times, as shown in figure \ref{fig:volume}(b). The experimental results are shown to match very well with the theoretical predictions at 39.4s, 29.4s and 9.4s before pinching occurs. The simulation is started at $t_d-t=394$ ms ($V=26$ mm$^3$). The simulation result at $t_d-t=200$ ms (after the simulation has been run for a physical time of 194 ms) is compared to the theoretical and experimental results. The theoretical, numerical and experiment curves all collapse 
perfectly, which again validates the present experimental and simulation approaches. 

 \subsection{Pinching and drop detachment}
 \label{sec:pinching}
\begin{figure}
\begin{center}
\includegraphics [width=1.0\columnwidth]{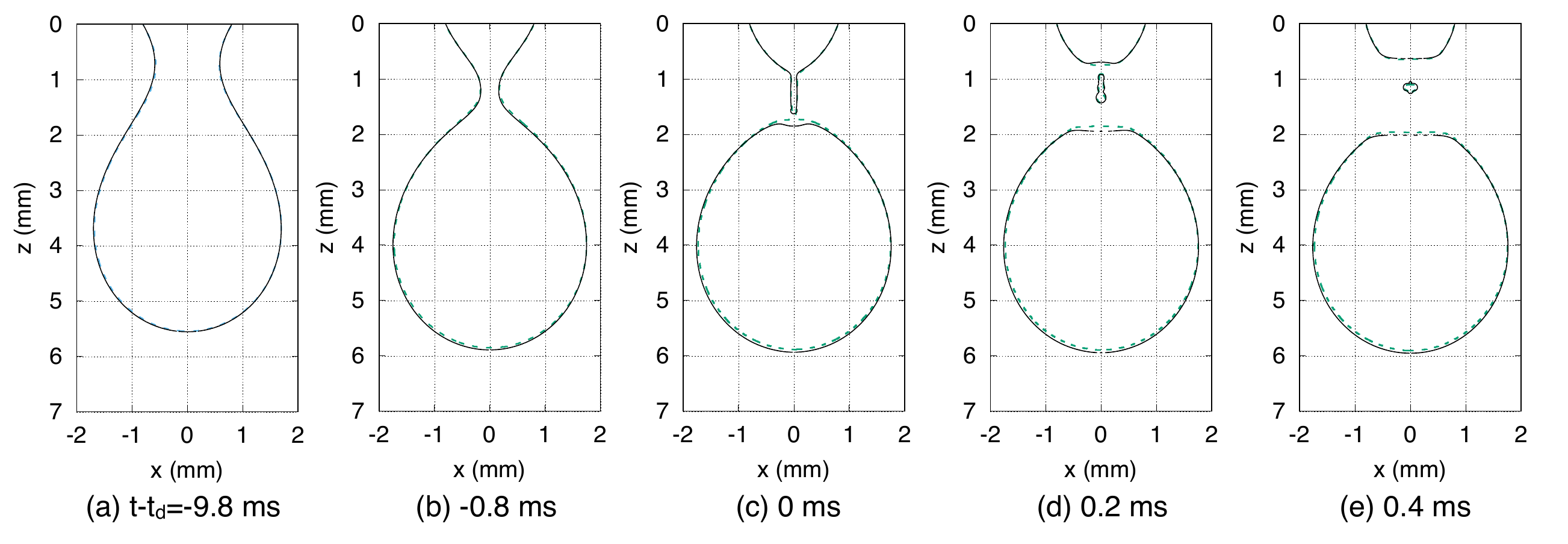}
\caption{Comparison between the numerical (solid lines) and experimental (dashed lines) results for the process of drop detachment.} 
\label{fig:dynamic}
\end{center}
\end{figure}

As the pendant drop reaches the critical volume, it becomes unstable. 
The interface evolution during the pinching process for both simulation and experiment is shown in figure \ref{fig:dynamic}. 
The numerical and experimental results generally agree very well for the formation of the neck and the liquid bridge, the detachment of the primary drop, and finally the formation of the secondary drop. 
In  figures \ref{fig:dynamic}(c)--(d), there exists a small discrepancy in the drop contours between experiment and simulation. This is due to the concave shape at the top of the drop, which cannot be seen from the experimental images taken from the lateral side. 

\begin{figure}
\begin{center}
\includegraphics [width=0.9\columnwidth]{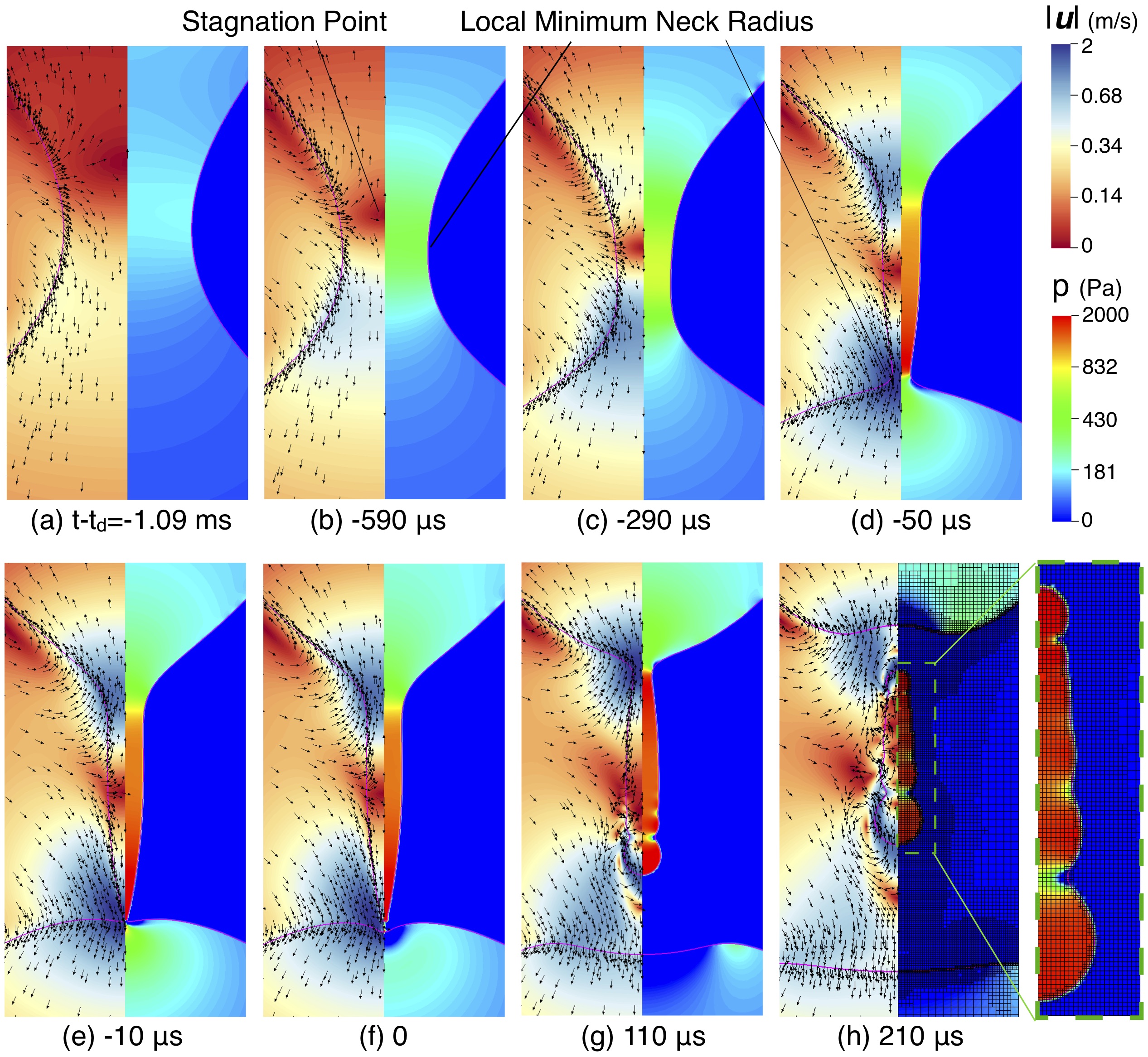}
\caption{Evolution of the velocity (left) and pressure (right) fields for the formation of primary and secondary drops. Skewed color scales have been used for better visualization. } 
\label{fig:pinching}
\end{center}
\end{figure}

To better elucidate the pinching dynamics and the formations of the primary and secondary drops, 
temporal evolutions of the pressure and velocity fields are plotted in figure \ref{fig:pinching}.
As the drop reaches the critical volume, a ``neck" develops between the nozzle and the pendant drop. The minimum radius of the neck ($x_{\min}$) decreases rapidly over time. As a consequence, the pressure in the neck, which is inversely proportional to the neck radius, also increases rapidly. The pressure difference between the neck and the regions above and below the neck expels the liquid away from the neck with increasing velocity, see figures \ref{fig:pinching}(a)--(d). The thinning process of the neck contributes to the elongation of the pendant drop and the neck turns into a thin liquid bridge. 
The minimum radius is initially located at about the center of the liquid bridge. The stagnation point is slightly higher than the location for the minimum neck radius.  As the liquid accelerates from the stagnation point toward the attached liquid and the primary drop, see \eg,  figure \ref{fig:pinching}(c),
the radius near the top and bottom of the bridge decreases faster than that near the center. 
The local radius minimum then shifts from the center to the bottom of the liquid bridge, 
see figure \ref{fig:pinching}(d). Pinching first occurs at the location for the new minimum radius 
near the bottom of the bridge, detaching the primary drop. After the pinch-off, the liquid filament rapidly retracts upward from the pinch-off location. Due to similar effect of the inertia of the upward fluid motion,
a new local minimum of radius develops at the top of the liquid bridge, see figure \ref{fig:pinching}(g), where soon another pinching happens. At the end, the liquid bridge is separated from the attached liquid and the primary drop, forming the secondary drop (see figure \ref{fig:pinching}(h)). A closeup of the secondary drop is also provided to show the high-resolution mesh used to resolve the pinching process. 
The dynamics of drop formation shown in the present experiment and simulation are consistent with former studies of drop formation \citep{Zhang_1995a, Wilkes_1999a, Popinet_2009a} and filament breakup \citep{Castrejon-Pita_2015a}.

\begin{figure}
\begin{center}
\includegraphics [width=0.99\columnwidth]{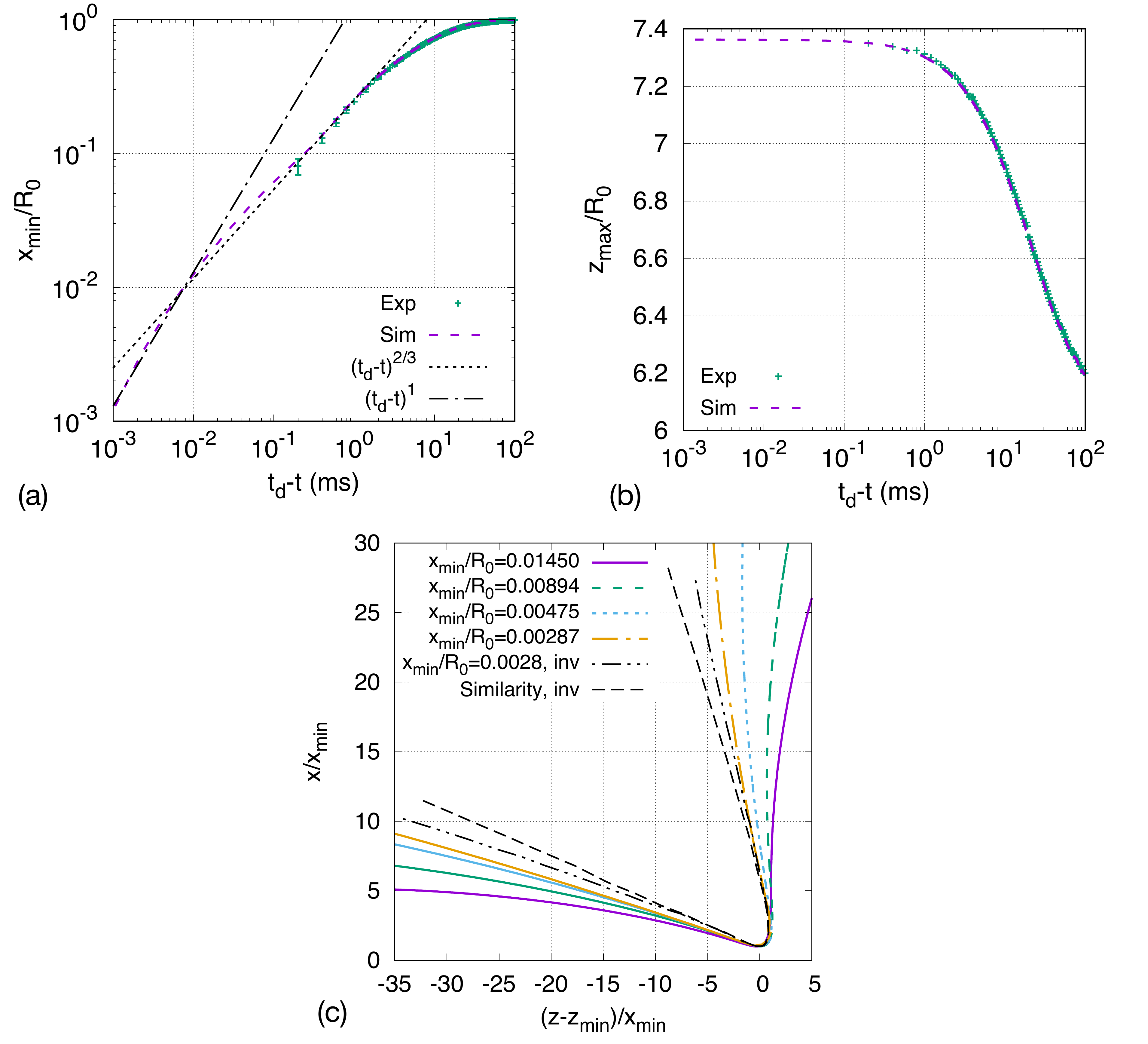}
\caption{Temporal evolution of (a) the minimum radius $x_{min}$, (b) the drop height $z_{max}$, and (c) the interface profiles near the pinching location prior to drop breakup. The \tcbl{dotted and dash-dotted} lines in (a) indicate the $(t_d-t)^{2/3}$ and $(t_d-t)$ power laws for the inertial and viscous regimes, respectively. The error bars on the experimental data in (b) are smaller than the line thickness and thus are not plotted. The simulation results shown in (c) approach the inviscid self-similar solution provided by \citet{Day_1998a}.} 
\label{fig:thread}
\end{center}
\end{figure}

\tcbl{
Since for the present problem $Oh\ll1$, the pinching process is mainly in the inertial regime where the temporal evolution of the minimum radius follows the 2/3 power law: $x_{\min}\sim (t_d-t)^{2/3}$. As the new minimum radius shifts from near the center toward the two ends of the liquid bridge, the downward flow from the neck to the primary drop slows down, reducing the local Reynolds number and bringing the pinching dynamics  into the viscous regime \citep{Castrejon-Pita_2015a}, where $x_{\min}\sim (t_d-t)$. 
The temporal evolution of $x_{\min}$ for both experiment and simulation is plotted in figure \ref{fig:thread}(a), where the two power-law scalings  and the transition from the inertial to viscous regimes can be clearly identified. As the viscous regime cannot sustain to the eventual breakup, another transition from the viscous regime to the inertial-viscous regime will occur in the pinching process at an even smaller time scale. Nevertheless, that time scale for the present problem with such a small $Oh$ is hard to resolve by simulation. Yet ignoring the inertial-viscous regime seems to introduce little effect on the formation of the primary drop. }

\tcbl{
The elongation of the drop due to the pinching process is measured and shown in figure \ref{fig:thread}(b). Again the numerical and experimental results agree very well. 
When the primary drop detaches from the liquid bridge, the drop height is about $z_{\max}/R_0=7.35$. Similar experiments of dripping water drops by \citet{Zhang_1995a} showed that $z_{\max}/R_0=9.92$ and 5.58 for nozzle radius $R_0=0.4$ and 1.6 mm, respectively. In the present study, $R_0=0.8$ mm, so the drop height at the detachment time is in a good agreement with the experimental results. 
}

\tcbl{Due to the low liquid viscosity in the present problem, when the liquid rushes from the neck toward the to-be-formed drop, the interface overturns before pinch-off occurs \citep{Day_1998a,Chen_2002a}. 
The overturning of the interface at the bottom of the liquid bridge can be identified 
with a careful look at figure \ref{fig:pinching}(e). 
A closeup of the interface near the pinch-off location is presented in figure \ref{fig:thread}(c) to better
show the overturning interface. 
The  simulation results are shown to approach the self-similar solution given by \citet{Day_1998a}. For the same minimum radius $x_{min}/R_0=0.0028$, the overturning interface obtained in the present simulation agrees well with the inviscid flow result \citep{Day_1998a}. 
}

The excellent agreement between the simulation, experiment, and theoretical results for both 
drop growth and detachment fully affirms that the drop formation is well captured 
and its effect on the subsequent fall of the drop has been faithfully incorporated in the present study. 

 \section{Results for shape oscillation}
 \label{sec:results_oscillation}
\subsection{Validation studies for oscillation and falling dynamics}
\begin{figure}
\begin{center}
\includegraphics [width=1.0\columnwidth]{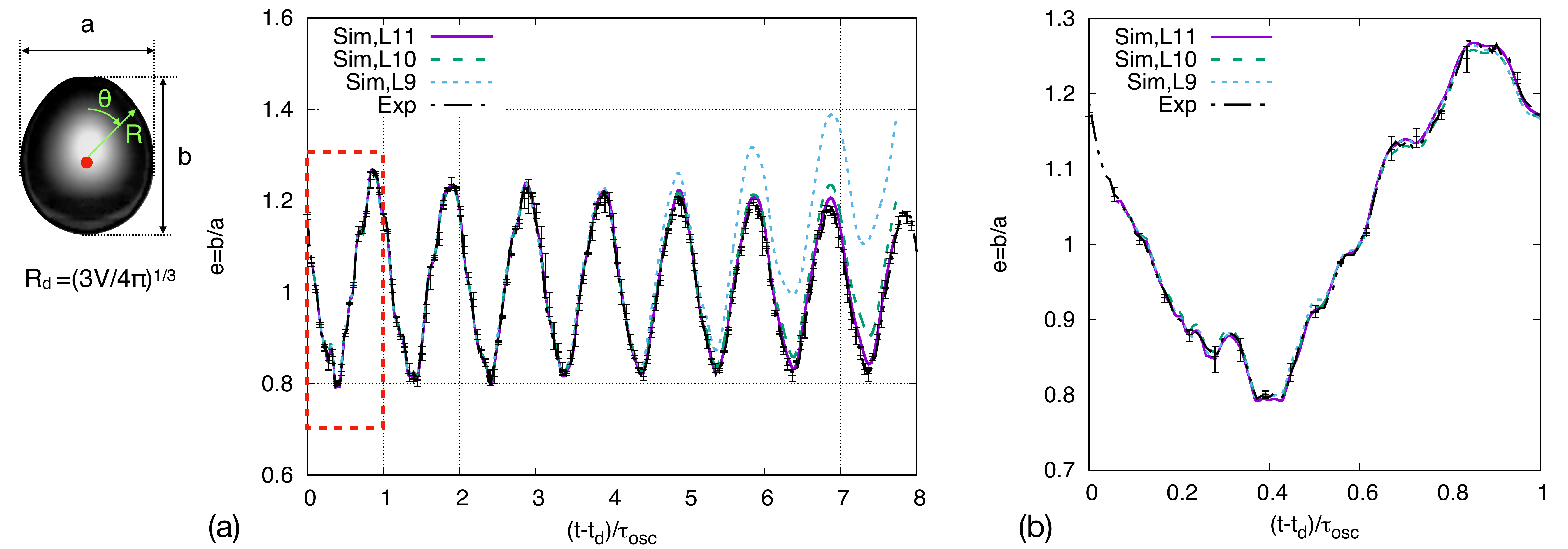}
\caption{Temporal evolution of the drop eccentricity for experiment and simulation. Here, the eccentricity is defined as $e = b/a$, where $b$ and $a$, as indicated, represent the height and width of the drop, respectively. The simulation results for different maximum mesh refinement levels ($L$) are compared to the experimental data in figure (a) and a closeup for $0<(t-t_d)/\tau_{osc}<1$ is given in figure (b). The corresponding minimum cell size $\Delta_{\min}$ for $L=11$, 10, and 9 are 3.12, 6.25, and 12.5 \textmu m, respectively. } 
\label{fig:oscillation}
\end{center}
\end{figure}

When the drop is detached from the nozzle, the drop shape is elongated and out of equilibrium. 
Under the action of surface tension, the drop starts to deform and oscillate. 
The eccentricity of the drop, defined as the ratio between the height ($b$) and the width ($a$) of the drop, $e=b/a$,  is a common parameter to characterize the shape deformation of an oscillating drop. \tcbl{The height $b$ is defined as the difference between the minimum and maximum z-coordinates of the droplet surface and thus does not account for the concave shape near the top of the drop shown in Fig.\ \ref{fig:dynamic}.} The temporal evolutions of $e$ obtained from simulations with different meshes are compared with the experimental measurement in figure  \ref{fig:oscillation}. It is observed that the second mode dominates the oscillation of $e$ and the time period agrees well with that for the second  mode of Lamb, $\tau_{2,Lamb}$. Therefore, $\tau_{2,Lamb}$ is taken to be the reference time scale for drop oscillation, 
namely $\tau_{osc}=\tau_{2,Lamb}$, and in figure \ref{fig:oscillation} time is normalized by $\tau_{osc}$. 
This indicates that the falling drop retains similar dominant frequency (or periods) as for a free drop. This observation is consistent with the former studies \citep{Lalanne_2013a, Staat_2017a, Bergeles_2018a}.
%Since the Lamb frequency is a good approximation for a falling drop in the short-term, Eq.\ \eqr{Lamb_freq} can be used to measure dynamic surface tension assuming the oscillation frequency is accurately measured by high-speed imaging \citep{Staat_2017a}. 

The angular frequency for the $n^{\mathrm{th}}$ spherical harmonic mode for small-amplitude oscillations of 
a free, viscous, and incompressible drop was derived by \citet{Lamb_1932a}, which is given as
\begin{align}
	\omega_{n,Lamb}^2 = \frac{(n-1)n(n+1)(n+2)\sigma}{[(n+1)\rho_l + n \rho_f] R_d^3 }\,. 
	\label{eq:Lamb_freq}
\end{align}
The frequency is $f_{n,Lamb}=\omega_{n,Lamb}/(2\pi)$ (for convenience $\omega$ is simply referred to as ``frequency" in the rest of the paper) and the time period is $\tau_{n,Lamb} = 1/f_{n,Lamb}=(2\pi)/\omega_{n,Lamb}$. 
For the second mode, the angular frequency is $\omega_{2,Lamb}=292$ s$^{-1}$, 
%the frequency $f_{2,Lamb}=46.6$ s$^{-1}$ 
and the oscillation period $\tau_{2,Lamb} = 21.5$ ms. 
The Lamb frequencies for other modes, $\omega_{n,Lamb}$, for the present drop size are listed in table \ref{tab:sph_mode}. 

The simulation results for all the three mesh refinement levels agree well with the experimental data 
at early times as shown in figure \ref{fig:oscillation}(a), though the results for the coarser meshes deviate from the experimental data at later times. For example, the curve for $L=9$ becomes different from the experimental data at about \tcbl{$(t-t_d)/\tau_{osc}>4.6 $}.
For the most refined case $L=11$ ($\Delta_{\min}\approx 3$ \textmu m  and $R_d/\Delta_{\min} \approx 595$), the numerical and experimental results match remarkably well in the time range ($(t-t_d)/\tau_{osc} \lesssim 8$) considered in the present study, indicating that $L=11$ is necessary and adequate to resolve the present problem. 

\begin{figure}
\begin{center}
\includegraphics [width=1.0\columnwidth]{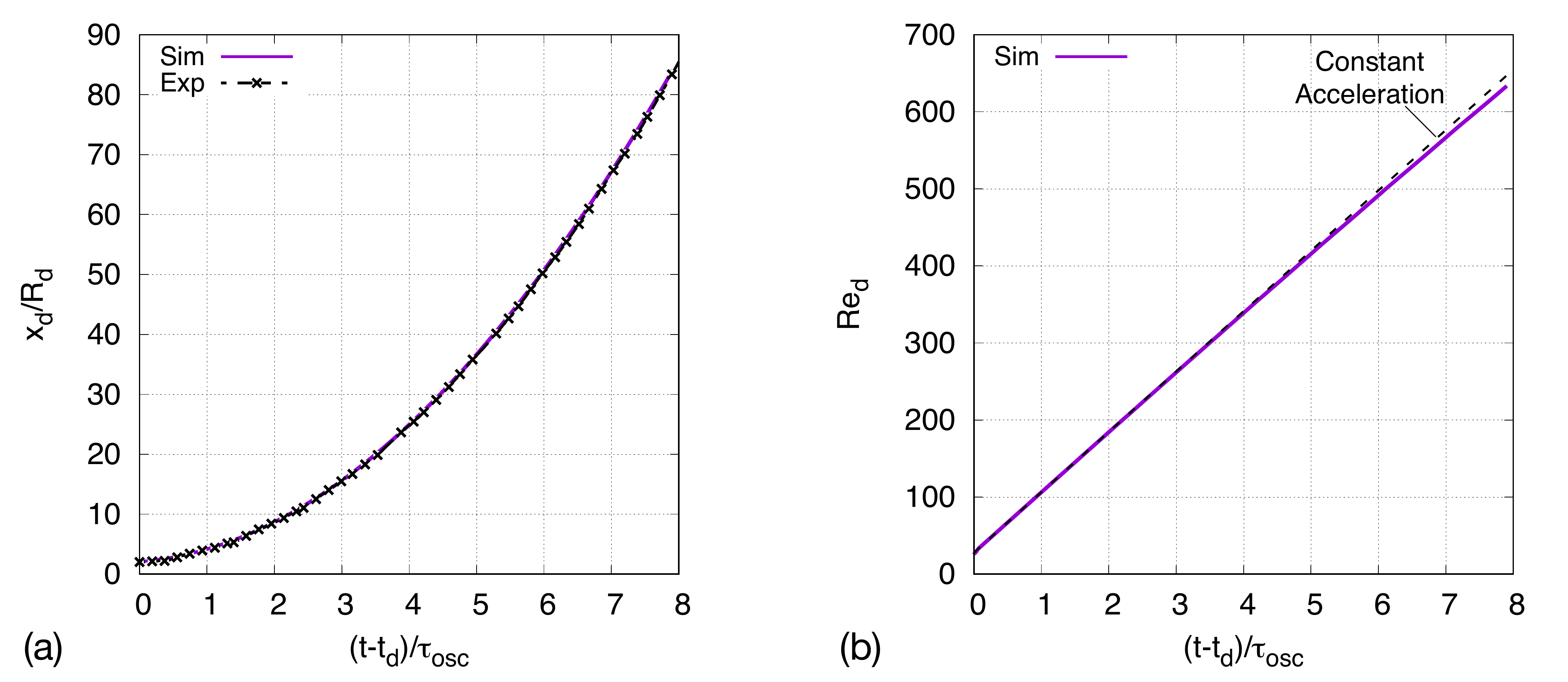}
\caption{Temporal evolutions of the drop centroid $x$-position and Reynolds number $Re_d$.} 
\label{fig:Reynolds}
\end{center}
\end{figure}

A closeup of the eccentricity evolution for $0<(t-t_d)/\tau_{osc}<1$ is presented in 
figure \ref{fig:oscillation}(b), from which it can be observed that the simulation results agree with experiment 
not only for the large-scale variation set by the dominant second mode, 
but also for the small-scale variations induced by the high-order oscillation modes. 
The temporal evolution of the drop centroid position is shown in figure \ref{fig:Reynolds}(a). 
The simulation and experiment results again match very well. Since the falling motion of the drop 
is coupled with the shape oscillation, the excellent agreement in high-level details 
between simulation and experiment for both eccentricity 
and drop trajectories fully validates the simulation results for both falling and oscillation dynamics 
of the drop. It also confirms that the axisymmetric approximation made in the present simulation is valid 
up to the time range considered. 

The evolution of the drop velocity, plotted in dimensionless form as the drop Reynolds number, is shown in figure \ref{fig:Reynolds}(b). A dashed line is given to indicate the evolution of  Re when the drop falls with no aerodynamic drag, namely undergoes a constant acceleration. In the short term, it is clear that the aerodynamic drag is small compared to the gravity force. The Reynolds number increases almost linearly, though small discrepancy can be identified for $(t-t_d)/\tau_{osc}>5$. The oscillation Reynolds number is $Re_{osc}=1/Oh_{osc}=360$. Initially $Re_d$ is smaller than $Re_{osc}$ but later overtakes and becomes larger than $Re_{osc}$. At $(t-t_d)/t_{osc}=7.9$, the drop Reynolds number, $Re_d=633$, is about 75\% larger than $Re_{osc}$. 
Nevertheless, it is observed that the dominant oscillation frequency for the falling drop is 
still well predicted by Lamb's linear theory for a free drop. 

\begin{table}
\centering
\begin{tabular}{lccccccccc} 
    \hline
n		&	2 & 3 & 4 & 5 & 6 & 7 & 8 & 9 & 10\\ [.5em]
$\omega_{n,Lamb}$ (s$^{-1}$)	& 	292.3 & 566.1 & 877.0 & 1223 & 1601 & 2009 & 2446 & 2909 & 3397\\ [.5em]
$\omega_{n,sim}$ (s$^{-1}$)	& 	306.8 & 552.2 & 859.0 & 1227 & 1595 & 2024 & 2454 & 2883 & 3436\\ [.5em]
$\beta_{n,Lamb}$ (s$^{-1}$)	& 	1.44 &  4.05 & 7.80 &  12.7 &  18.8 &  26.0 &  34.4 &  43.9 &  54.6 \\[.5em]
$A_{n,0}$					& 	0.10   &    0.037  & 0.022 &  0.012 &  0.0066 &  0.0031 &  0.0011  & -0.00045  & -0.0014\\[.5em]
$\alpha_{n,0}$				& 	0.144  &   0.0672  & 0.0393 &  0.0280 &  0.0210 &  0.0165 &  0.0135  & 0.0110  & 0.0092\\[.5em]
$\phi_{n}/\tau_2$			& 	0.12   &    0.08      & 0.06     &  0.04     &  0.031   &  0.024   &  0.020    & 0.017    & 0.015\\
%$\beta_{n,sim-peak}$ (s$^{-1}$)	& 	1.35 &  4.12 & 3.62 &  13.7 &  18.0 &  27.1 &  26.6 &  41.2 &  53.0 \\[.5em]
%$\beta_{n,sim-valley}$ (s$^{-1}$)	& 	1.67 &  3.45 & 11.4 &  13.0 &  25.8 &  26.5 &  32.1 &  40.2 &  48.6 \\[.5em]
$\beta_{n,peak}$ (s$^{-1}$)	& 	1.35 &  4.12 & - & - & - & - & - & - & -   \\[.5em]
$\beta_{n,valley}$ (s$^{-1}$)	& 	1.67 &  3.45  & - & - & - & - & - & - & - \\[.5em]
$\alpha_{n,0,peak}$			& 	0.148 & 0.0673	& - & - & - & - & - & - & -   \\[.5em]
$\alpha_{n,0,valley}$			& 	0.141 & 0.0652 & - & - & - & - & - & - & -  \\
    \hline
\end{tabular}
\caption{Results for the spherical harmonic mode analysis for the oscillation of the falling drop.
\tcbl{The frequency $\omega_{n,Lamb}$ and damping rate $\beta_{n,Lamb}$ are calculated following the linear theory of \citet{Lamb_1932a}. 
The primary frequency $\omega_{n,sim}$ is measured through the frequency spectrum of the computed Fourier-Legendre coefficients $A_n$. The value of $A_n$ at $t=t_d$ is denoted by $A_{n,0}$, while the amplitude ($\alpha_n$) of the oscillation of $A_n$ at $t=t_d$ is represented by $\alpha_{n,0}$. The initial phase of the oscillation of $A_n$ is denoted by $\phi_{n}$. The values of $A_{n,0}$, $\alpha_{n,0}$ and $\phi_n$ are obtained from simulation results for drop formation. Exponential functions are used to fit the peaks and valleys of the temporal evolution of $A_n$ for $n=2,3$. The fitted initial oscillation amplitudes and damping rates for peaks and valleys are represented by $\alpha_{n,0, peak}$ and $\alpha_{n,0,valley}$, and $\beta_{n,0, peak}$ and $\beta_{n,0,valley}$, respectively. }}
\label{tab:sph_mode}
\end{table}
  
 \subsection{Spherical harmonic mode decomposition}
To better understand 
the shape oscillation of the falling drop, the instantaneous shape of the drop is decomposed 
into spherical harmonic modes \citep{Basaran_1992a, Lalanne_2013a}. 
\tcbl{The temporal 
evolution and frequency spectra of the mode amplitudes will be presented to analyze the effects of the drop formation, 
the nonlinear dynamics, and the falling motion on the shape oscillation. }

\begin{figure}
\begin{center}
\includegraphics [width=0.98\columnwidth]{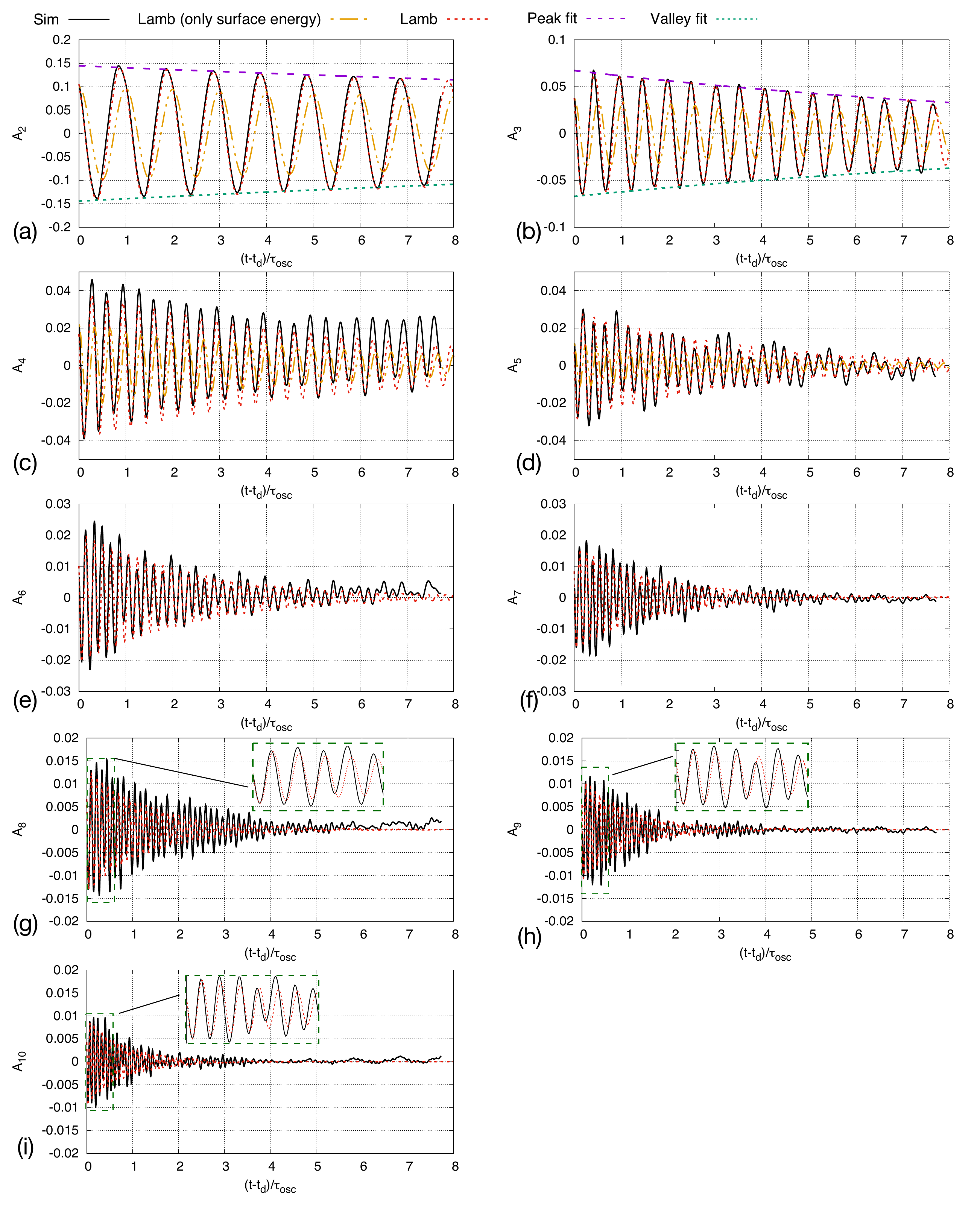}
\caption{Temporal evolutions of the Fourier-Legendre coefficients, $A_n$, for different spherical harmonic modes, comparing the simulation results with the linear free-drop model based on the theory of \citet{Lamb_1932a}, with and without the initial kinetic energy. The exponential decay of the oscillation amplitudes for the peaks and valleys are also indicated in (a) and (b) for $n=2$ and 3 modes.  } 
\label{fig:mode}
\end{center}
\end{figure}

The shape of an axisymmetric drop can be described by the radius of the drop contour with respect to the centroid $R$
as a function of the colatitude $\theta$ (which is taken to be zero at the top of the drop), as shown in figure \ref{fig:oscillation}. 
For an oscillating drop, $R=R(\theta,t)$, and can be expanded as the superposition of 
spherical harmonic modes as 
\tcbl{
 \begin{align}
 	\frac{R(\theta,t)}{R_d} = \sum_{n=0}^{\infty}A_n(t) P_n(\cos (\theta))\, ,
 \end{align}
where $P_n$ is the Legendre polynomial of degree $n$ and $A_n$ is the corresponding Fourier-Legendre coefficient, which represents the amplitude of the $n^{\mathrm{th}}$ spherical harmonic mode. 
Assuming incompressibility,  the drop volume is fixed and $A_0=1$. 
Furthermore, for the analysis of the falling drop, a reference frame moving with the drop velocity is used and the origin is set as the centroid of the drop. As a result, $A_1=0$. }
 The temporal evolutions of $A_2$ to $A_{10}$ for the simulation results are shown in figure \ref{fig:mode}. 
\tcbl{A grid refinement study has been performed to confirm that the results presented are mesh independent, see appendix \ref{sec:mode_grid}. }
 
The Fourier-Legendre coefficients at $t=t_d$  are denoted as $A_{n,0}$ 
and the values are listed in table \ref{tab:sph_mode}. 
The initial amplitudes for spherical harmonic modes generally decrease with the mode number $n$. 
The amplitudes of higher order modes ($n>2$) are finite and cannot be ignored. 
For example, $A_{5,0}$ and $A_{7,0}$ are about 11\% and 3\% of $A_{2,0}$. 
The small-scale spatial variations in the drop contours near the top of the drop (see figure \ref{fig:dynamic}(c)), 
which are in turn induced by the pinching process, 
contribute to the finite amplitudes of the high order oscillation modes. 

\begin{figure}
\begin{center}
\includegraphics [width=0.99\columnwidth]{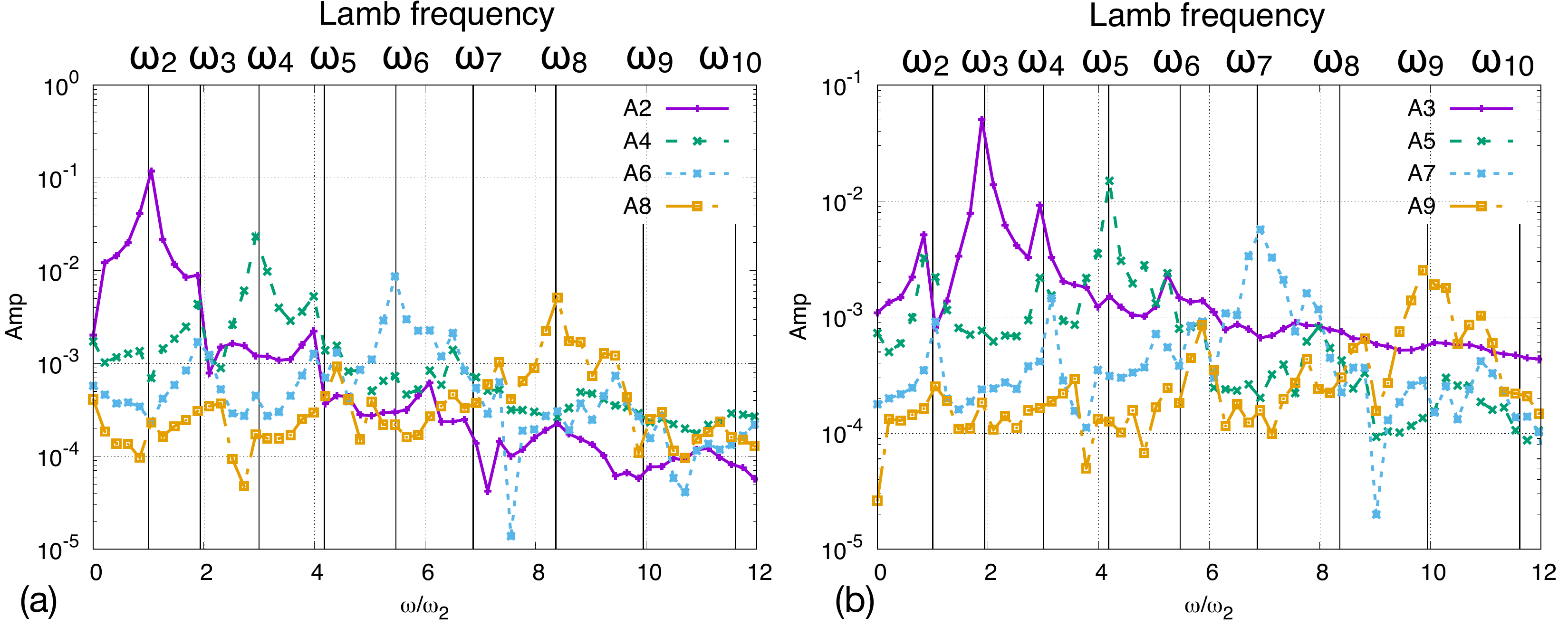}
\caption{Frequency spectra of Fourier-Legendre coefficients for (a) even and (b) odd spherical harmonic modes, indicating the effect of mode coupling.} 
\label{fig:mode_coupling}
\end{center}
\end{figure}

The frequency spectra of  $A_n$ are shown in figure \ref{fig:mode_coupling}, 
from which the primary frequency for each mode can be identified. 
The values of the primary frequencies for simulation, $\omega_{n,sim}$, 
are given in table \ref{tab:sph_mode}. 
It can be seen that the oscillation frequency agrees well with the Lamb frequency. 
This conclusion is valid not only for the dominant $n=2$ mode 
(as already shown in figure \ref{fig:oscillation}) but also for other modes up to $n=10$. 
It can be observed from figure \ref{fig:mode} that, at the end of the simulation, $(t-t_d)/t_{osc}\approx7.9$, 
the drop Reynolds number, $Re_d=633$, is about 75\% larger than $Re_{osc}$, 
yet the agreement with the Lamb frequency is still very good. 

According to the nonlinear analysis of \citet{Tsamopoulos_1983a}, the leading term in the decrease 
of oscillation frequency due to finite amplitude is second order. For the dominant second mode, 
the initial amplitude $A_{2,0}$ is about 10\%. The correction of frequency due to nonlinear effects is about 1\%, which is quite small. This explains why the linear theory of \citet{Lamb_1932a} remains a very good
 approximation for present case, even though the mode amplitudes are finite. 
%Nevertheless, though the nonlinear dynamics has little influence on the dominant oscillation frequency, as will be shown later, 
%it does have an impact on other aspects of oscillation dynamics. 
%The post-formation state of the drop does trigger a moderately nonlinear oscillation, which (as will be shown later) interacts with the falling motion and induce a complex transient flow around the drop. 

\subsection{Linear oscillation of a free viscous drop}
The short-term oscillation of the drop is mainly controlled by the capillary effect, however, it is also significantly affected \tcbl{by} the drop formation, the nonlinear dynamics due to finite oscillation amplitudes, and the falling motion. To better understand these effects on the oscillation dynamics, the simulation results are 
compared to the linear theory of  \citet{Lamb_1932a} for the linear oscillation of a free viscous drop. 

The Fourier-Legendre coefficients for the n$^{th}$ Lamb mode, $A_{n,Lamb}$, are given  as
 \begin{align}
 	A_{n,{Lamb}}(t) = \alpha_n \cos[ \omega_{n,Lamb}(t +\phi_{n})] \, .
	\label{eq:mode_amp}
 \end{align}
For a viscous drop, the oscillation amplitude $\alpha_n$ decreases in time due to viscous dissipation. 
For small $Oh_{osc}$, the viscous damping effect causes an exponential decay of $\alpha_{n}$, 
 \begin{align}
 	\alpha_n(t) = \alpha_{n,0} \exp(-\beta_{n,Lamb} t) \, ,
	\label{eq:damping}
 \end{align}
where \tcbl{$\beta_{n,Lamb}$ is the damping rate, given by \cite{Lamb_1932a} as} 
 \begin{align}
 	\beta_{n,Lamb} = \frac{(n-1)(2n+1)\nu_l}{R_d^2}\, . 
	\label{eq:damping_Lamb}
 \end{align}
Then Eq.\ \eqr{mode_amp} can be rewritten as 
 \begin{align}
 	A_{n,{Lamb}}(t) = \alpha_{n,0} \exp(-\beta_{n,Lamb} t)  \cos[ \omega_{n,Lamb}(t +\phi_{n})] \, .
	\label{eq:Lamb_model}
 \end{align}

The viscous damping influences the oscillation frequency as   
$\omega_{n}^{*2} = {\omega_{n,Lamb}^{2} - \beta_{n,Lamb}^2}$. For the present problem $\beta_{n,Lamb} \ll \omega_{n,Lamb}$ (see table \ref{tab:sph_mode}),
as a result, the decrease of frequency due to viscous effect is negligible. This also explains why 
the dominant oscillation frequency agrees so well with the Lamb frequency (Eq.\ \eqr{Lamb_freq})
as already shown in figure \ref{fig:oscillation}. 

In Eq.\ \eqr{Lamb_model}, there are in total four parameters, $\omega_{n,Lamb},\beta_{n,Lamb},\alpha_{n,0}, \phi_{n}$. The frequency $\omega_{n,Lamb}$ and damping rate $\beta_{n,Lamb}$, as shown in Eqs.\ \eqr{Lamb_freq} and \eqr{damping}, depend only on the fluid properties. In contrast, 
the initial oscillation amplitude of the Fourier-Legendre coefficient, $\alpha_{n,0}$, and is the initial phase, $\phi_{n}$, 
are determined by the drop formation process and the resultant post-formation state, including both the shape (surface energy) and the velocity field (kinetic energy). 
%The values of $\alpha_{n,0}$ and $\phi_{n}$ were
%obtained by fitting the simulation results. 

\tcbl{
\subsection{Effect of the initial kinetic energy in the drop}
%key message: a_n,0>A_n,0, phi_n/=0, cannot be ignored
Conventionally, the surface energy contained in the initial shape is assumed to dominate 
the initial state of drop oscillation and the initial kinetic energy (velocity field) is usually ignored. 
The present study that covers both the drop formation and subsequent oscillation provides 
an opportunity to reexamine this assumption. }

If the kinetic energy in the initial condition is ignored, \ie, the velocity field is zero everywhere, or a static drop
with the same shape as the post-formation drop is released, then $\alpha_{n,0}=A_{n,0}$ and $\phi_n=0$
and Eq.\ \eqr{Lamb_model} becomes 
 \begin{align}
 	A_{n,{Lamb,surf}}(t) = A_{n,0} \exp(-\beta_{n,Lamb} t)  \cos[ \omega_{n,Lamb}(t)] \, .
	\label{eq:Lamb_model_shapeOnly}
 \end{align}
The results of Eq.\ \eqr{Lamb_model_shapeOnly} for the first four modes ($n=2$ to 5)
are plotted in figure \ref{fig:mode}. It is clear that the model including only the surface energy 
in the initial state yields results that are very different from  the simulation results, 
even though the Fourier-Legendre coefficients for the exact initial shape of the drop, $A_{n,0}$, 
have been used. A close examination of figure \ref{fig:mode}(a) indicates that 
the deviation starts right at $t-t_d=0$. The computed $A_2$ decreases faster and to a lower minimum 
than that predicted by the model. 
The decrease of $A_2$ represents that the drop deforms from the prolate (elongated) to the oblate (flattened) shapes. 
Therefore, the drop in simulation is flattened faster and to a larger extent compared to the model prediction. 
The discrepancy is due to the remaining effect of pinching dynamics and 
the non-uniformly distributed kinetic energy in the post-formation drop. 
As discussed above in section \ref{sec:pinching}, the high pressure in the liquid bridge expels 
fluid toward the drop (which even induces overturning of the interface at the top of drop). 
As a consequence, when the drop is just detached from the liquid bridge, 
the top portion of the drop retains a significant downward velocity, 
which contribute to strengthening the prolate-to-oblate deformation, in addition to the capillary effect.  
The results clearly lead to the conclusion that the initial kinetic energy is as important as the 
initial surface energy to the shape oscillation  and should not be ignored. 

\tcbl{The key contributions of the initial kinetic energy to the shape oscillation are the amplification of
 $\alpha_{n,0}$ and the non-zero initial phase angle $\phi_{n}$. 
The values of $\alpha_{n,0}$ and $\phi_{n}$ for different modes can be obtained by 
fitting Eq.\ \eqr{Lamb_model} with the simulation results near $t-t_d=0$.
As shown in table \ref{tab:sph_mode}, $\alpha_{n,0}>|A_{n,0}|$ and $\phi_{n}\neq 0$
are true for all the modes considered here. The amplification of $\alpha_{n,0}$ and the non-zero 
initial phase angle due to drop formation were also observed in the experiments of \citet{Becker_1991a}, 
though the physics behind them was not discussed. 
With the corrected $\alpha_{n,0}$ and $\phi_{n}$, Eq.\ \eqr{Lamb_model} 
yields a much better agreement with the simulation results for the whole time range considered, 
see figure \ref{fig:mode}. (Hereafter, Eq.\ \eqr{Lamb_model} with corrected values 
of $\alpha_{n,0}$ and $\phi_{n}$ is referred to as the linear free-drop model.)
Considering the fact that the linear free-drop model still ignores the effects of falling motion and  
nonlinear dynamics, the agreement between the model and the simulation is quite impressive
for the $n=2$ and 3 modes. }

In spite of the apparent good agreement between the linear free-drop model and the simulation
results for the lower-order modes ($n=2,3$), significant differences exist in the higher-order modes ($n \ge 4$). 
At early time (\tcbl{$t\lesssim 5\tau_{osc}$}) the falling velocity is small and thus the effect of the falling motion is negligible, 
the discrepancy is thus mainly due to the nonlinear effects, which are in turn triggered by 
the finite mode-amplitudes when the drop is formed. 
At later time, when the drop velocity becomes large, $Re_d > Re_{osc}$, 
the contribution of the falling motion to the discrepancy becomes significant. 
These two effects are discussed in sequence in the following sections. 
%The ratio $Re_d/Re_{osc}$ can be a good parameter to characterize the relative importance of the effect of falling motion compared to the capillary effect on drop oscillation, yet parametric studies for different $Oh$ and longer simulations will be required to identify the range of $Re_d/Re_{osc}$ during which the oscillation frequency for a falling drop remains in agreement  with the Lamb frequency. 

\subsection{\tcbl{Effect of mode coupling and energy transfer}}
As summarized by \citet{Becker_1991a}, typical nonlinear effects 
in shape oscillation include a) the dependence of the oscillation frequency on the amplitude, 
b) the asymmetry of the oscillation amplitude, and c) the coupling between modes. 
As shown in figure \ref{fig:oscillation} the variation in frequency is small for the present case, 
however, the other two nonlinear effects can be clearly identified. 

A close look at figure \ref{fig:mode} shows that 
the oscillation amplitude of $A_n$ is generally asymmetric, 
namely the oscillation amplitudes corresponding to the peaks and valleys are different. 
The asymmetry of oscillation amplitude is most profound for the $n=4$ mode: 
the temporal evolution of $A_4$ is clearly shifted upward, see figure \ref{fig:mode}(c). 
(A physical explanation for the strong nonlinear effect for the $n=4$ mode is 
to be given later.)
Similar but less obvious upward shifting in the mode amplitude evolution can also be identified 
for the $n=6$ and 8 modes. 
The asymmetric behavior is less obvious for the lower-order modes ($n=2$ and 3). 
To better illustrate the asymmetric behavior, 
the exponential function  (Eq.\ \eqr{damping}) is used to fit the peaks and valleys
of the temporal evolutions of $A_2$ and $A_3$. 
The fitted initial amplitudes and damping rates for the peaks and valleys are different 
as shown in table \ref{tab:sph_mode}. It is shown that $\alpha_{n,0,peak}>\alpha_{n,0,valley}$ for both 
$n=2$ and 3 modes. For the damping rate, $\beta_{2,peak}<\beta_{2,valley}$ 
while $\beta_{3,peak}>\beta_{3,valley}$. The damping rate prediction of Lamb (Eq.\ \eqr{damping_Lamb}) 
lies in between the damping rates for the peaks and valleys. 
Due to the strong non-monotonicity in the decay of the oscillation amplitude for the higher-order modes, 
it is infeasible to fit the amplitude with an exponential function. 

Another important nonlinear effect on drop oscillation is the interaction between 
different spherical harmonic modes through energy transfer. 
When energy is added or extracted from a specific mode, the oscillation amplitude of that mode 
will be amplified or suppressed, respectively. As a result, the decay of oscillation amplitude 
becomes non-monotonic, see figures \ref{fig:mode} (d--i).  
It is conventionally considered that the nonlinear effects arise due to 
a large amplitude, however, it is observed here that the nonlinear effect is stronger 
for the higher-order modes ($n\ge 4$) than the lower-order modes ($n=2,3$)
while the amplitudes of the former are actually smaller than of the latter. 
This interesting behavior has also been observed in experiments
and can be explained through mode coupling \citep{Becker_1991a}. 
For the present problem, the energy stored in the lower-order modes is significantly 
larger than that in the higher-order modes, 
see $\alpha_{n,0}$ values in table \ref{tab:sph_mode}. 
Therefore, when a small energy transfer between the lower-order and higher-order modes, 
its effect on the lower-order mode amplitude is small but it can modify the higher-order mode amplitude 
significantly. 

\tcbl{
Due to the large water-to-air density ratio in the present problem, the Lamb frequency is similar to 
the Rayleigh frequency 
\begin{align}
	\omega_{n,Rayleigh}^2 = \frac{(n-1)n(n+2)\sigma}{\rho_l R_d^3 }\,. 
	\label{eq:Rayleigh_freq}
\end{align}
An important feature of the Rayleigh frequency is that $\omega_2$ and $\omega_4$ are commensurate ($\omega_4=3\omega_2$), see table \ref{tab:sph_mode}. 
As a result, there exist a resonant effect in the coupling between the $n=2$ and $n=4$ modes \citep{Tsamopoulos_1983a,Natarajan_1987a}. As the $n=2$  mode is the dominant mode that 
contains the most of the oscillation energy, the $n=4$ mode is modulated significantly due to 
the resonant energy transfer between the two modes. This explains why the nonlinear effect is always 
the most intense for the $n=4$ mode. 
}

\tcbl{
The effect of mode coupling is also shown in the frequency spectra of the spherical harmonic mode amplitudes $A_n$, 
see figure \ref{fig:mode_coupling}. 
While the linear free-drop model yields a single frequency for each mode, 
$\omega_{n,Lamb}$ (indicated by the vertical lines), 
the spectra of computed $A_n$ show multiple frequencies for modes $n>2$.
For the fundamental $n=2$ mode, only the primary frequency $\omega_2$ is observed. 
(Other smaller peaks in the $A_2$ spectrum correspond to the even number times of the primary frequency, 
such as $2\omega_2$, $4\omega_2$, $6\omega_2$.)
For a given mode $n>2$, the spectrum shows a primary frequency that agrees well with $\omega_{n,Lamb}$,  
and also multiple secondary frequencies corresponding to other modes ($\omega_m$ with $m\ne n$) which  
interact with the $n^{th}$ mode. 
In the spectrum of $A_4$, secondary frequencies $2\omega_2$ and $4\omega_2$ are observed, 
(note the small difference between $4\omega_2$ and $\omega_5$,)
which is another evident for its strong coupling with the $n=2$ mode.
A  close look indicates that  the $A_6$ spectrum also shows similar secondary frequencies at  $2\omega_2$ and $4\omega_2$. 
Former studies  have shown that an initial second-mode deformation will excite
even modes due to mode coupling \citep{Tsamopoulos_1983a, Basaran_1992a}. 
Therefore, though the coupling between the dominant $n=2$ mode and other higher-order even modes 
like $n=6$ is not as strong as with the $n=4$ mode, their spectra also show the influence from the second mode. 
}

Furthermore, a drop with initial finite-amplitude deformation of odd modes will transfer energy to the fundamental $n=2$ mode
and excite the oscillation of the latter \citep{Basaran_1992a}. In figure \ref{fig:mode_coupling}, a secondary frequency of $\omega_2$ is observed 
in the spectra of the odd modes $n=3,5,7,9$. Due to the resonant coupling between the $n=2$ and 4 modes, the oscillation 
energy from the odd modes can also be transfered to the $n=4$ mode through the intermediary $n=2$ mode. As results, 
the spectra of the odd modes also show a secondary frequency $\omega_4$. 
Finally, another commensurate relation exists 
between the $n=5$ and 8 modes, namely $\omega_8=2\omega_5$, and therefore, 
there exists a resonant coupling between the two. That explains why a secondary frequency $\omega_8$ arises 
in the spectrum of $A_5$. It can be seen from figure \ref{fig:mode} that, although the decay in oscillation amplitude is non-monotonic due to mode-coupling, the viscous damping rates 
are generally consistent with Lamb's prediction (see the results for the linear free-drop model). 
However, the $n=8$ mode seems to be an exception, the decay of oscillation amplitude 
is slower than the linear free-drop model, which is due to the resonant coupling between the $n=5$ and 8 modes. 

\subsection{Effect of falling motion}
The effect of the drop fall  on the shape oscillation is initially small, 
yet as the drop falling velocity increases in time, its impact on drop oscillation is enhanced. 
The influence of the falling motion on the shape oscillation can be identified 
through the asymmetric oscillation amplitude. 
The asymmetric amplitude in $A_n$ (such as $n=4$) for $(t-t_d)/\tau_{osc}\lesssim 4$ 
is due to the nonlinear effect. 
If the drop does not fall, then as the oscillation amplitude decreases with time, the nonlinear effect will become weaker 
and the level of asymmetry will also decrease over time. For the falling drop considered here, 
it is observed in figure \ref{fig:mode}(c) that the difference between the peak and valley amplitudes decreases 
initially but then remains at a similar level for $(t-t_d)/\tau_{osc}\gtrsim 4$, which is due to the interaction between 
the drop and the external flow induced by the falling motion. 
In the long term when the drop reaches its terminal falling velocity, (the drop can reach a fixed shape \citep{Feng_2010a}
or still oscillate \citep{Helenbrook_2002a} depending on $Re_{d,\infty}$ and $We_{d,\infty}$),
the balance between surface tension and shear stress induced by the external flow results in 
a non-spherical equilibrium drop shape which exhibit non-zero mode amplitudes ($A_{n,eq}\ne 0$). 
Although the time period considered here is far from the equilibrium state, 
the shear stress induced by falling motion already has an impact on the drop shape 
and enhance the asymmetry in oscillation amplitudes. 
The asymmetric effect is reflected as an upward shit of $A_n$ for the higher-order even modes ($n=4,6,8,10$)
and is negligibly small for higher-order odd modes ($n=5,7,9$). 
For the lower-order modes ($n=2,3$), the oscillation amplitudes are large and thus the capillary effect dominates. Therefore, the effect of falling motion is less profound.  

There also exists an energy transfer between the falling motion and the shape oscillation. It can be observed from figure 
\ref{fig:mode}(c) that for $(t-t_d)/\tau_{osc}\gtrsim 4$, the oscillation amplitude decays much slower than the 
linear free-drop model. The energy dissipated by viscosity is compensated by the energy from the falling motion. 
Similar slower decay in oscillation amplitude for $(t-t_d)/\tau_{osc}\gtrsim 4$ can also be observed 
in figures \ref{fig:mode}(d--i) for other high-order modes. 

%continue here. 
\section{Results for the transient flow field }
 \label{sec:results_transient}
\begin{figure}
\begin{center}
\includegraphics [width=0.75\columnwidth]{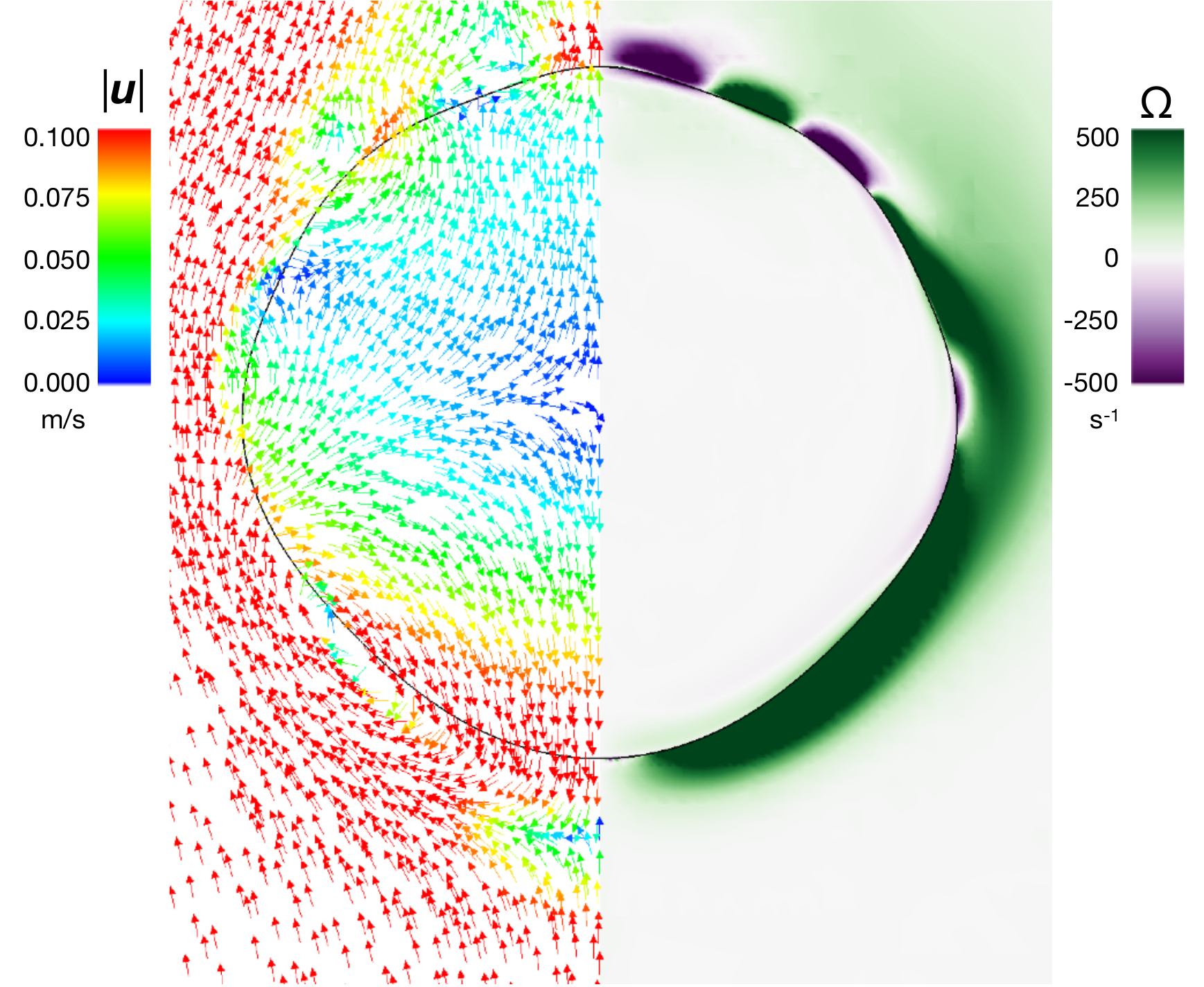}
\caption{\tcbl{Simulation results for }the velocity (left) and vorticity (right) fields around the drop at $(t-t_d)/\tau_{osc}=0.56$. } 
\label{fig:multimode}
\end{center}
\end{figure}

 The multi-mode oscillation of the falling drop is accompanied by a complex transient velocity field around the drop, 
see figure \ref{fig:multimode}. The snapshot shown here is taken soon after the drop is formed, at $(t-t_d)/\tau_{osc}=0.5$, \tcbl{from the simulation results}. 
The inward and outward motions of the interface can be observed from the velocity vector field. 
The oscillating motion of the interface induces swirling motion of the fluid near the drop, which can be visualized by the vorticity ($\Omega$) field, 
as shown in the right half of figure \ref{fig:multimode}. 

\tcbl{
\subsection{Asymptotic limits}
}
To better understand the development of the flow field for the falling and oscillating drop, we first look at the two 
asymptotic limits: 1) the case when the drop is freely oscillating but not falling, and 2) the case when the 
drop is falling but without oscillation. 

\begin{figure}
\begin{center}
\includegraphics [width=0.9\columnwidth]{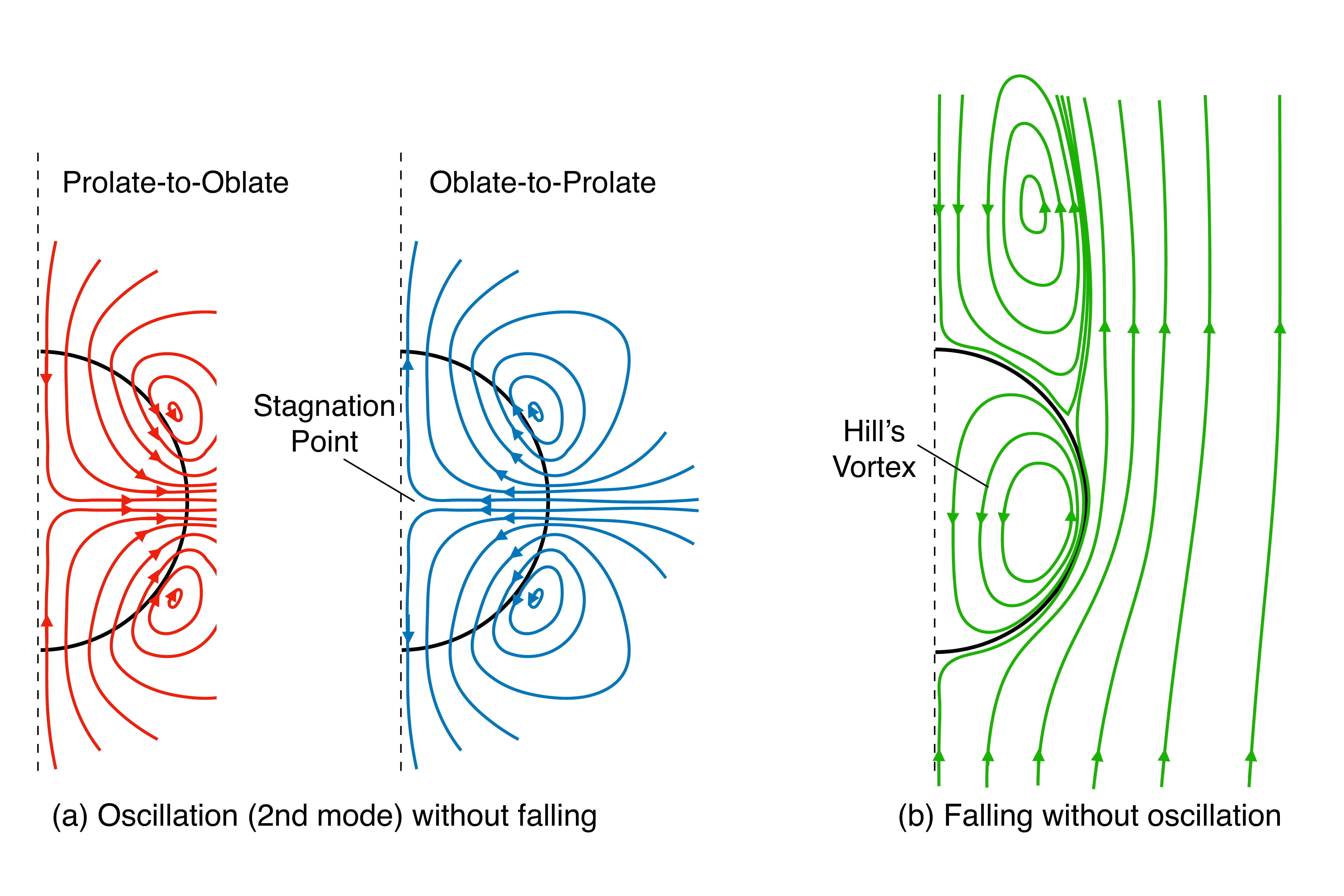}
\caption{Schematics of the flow field for (a) a drop that is oscillating without falling motion, and (b) a drop that is falling without oscillation. Figure (a) is adapted from our simulation of a free drop undergoing only second mode oscillation. In figure (b) the streamlines are sketched based on the simulation results by \citet{Feng_2010a} for $Re_d=200$ and $We_d=1$.} 
\label{fig:streamline_sketch}
\end{center}
\end{figure}

A representative flow field around an freely oscillating drop is shown in figure \ref{fig:streamline_sketch}(a). 
The simple case shown here contains only the second mode. As a response to the oscillation, two vortices are formed outside the drop with opposite rotation directions. The directions of the two vortices change within the oscillation cycle. 
When higher-order modes exist, more vortices will arise as can be seen in figure \ref{fig:multimode}. 

As the drop falls, it accelerates  and the relative velocity between the drop and the surrounding air increases in time 
until the terminal velocity is reached. When the drop Reynolds and Weber numbers are small, the drop will eventually reach a steady state. 
For this limiting case where the drop is falling without oscillation, the internal flow pattern is dictated by the external shear. 
In the Stokes limit, the drop shape will remain spherical and the flow circulation inside the drop 
is known as Hill vortex \citep{Hill_1894a}. For finite but small Reynolds and Weber numbers, 
the drop will not be perfectly spherical but the internal flow remains similar to Hill vortex \citep{Feng_2010a}. 
A representative flow field for a falling drop without oscillation is shown in figure \ref{fig:streamline_sketch} (b), 
which is sketched based on the simulation results of \citet{Feng_2010a} \tcbl{for $Re_d=200$ and $We_d=1$. There exist only one vortex, similar to the Hill vortex, inside the drop.} 

\begin{figure}
\begin{center}
\includegraphics [trim=0cm 0 0cm 0,clip, width=1.0\columnwidth]{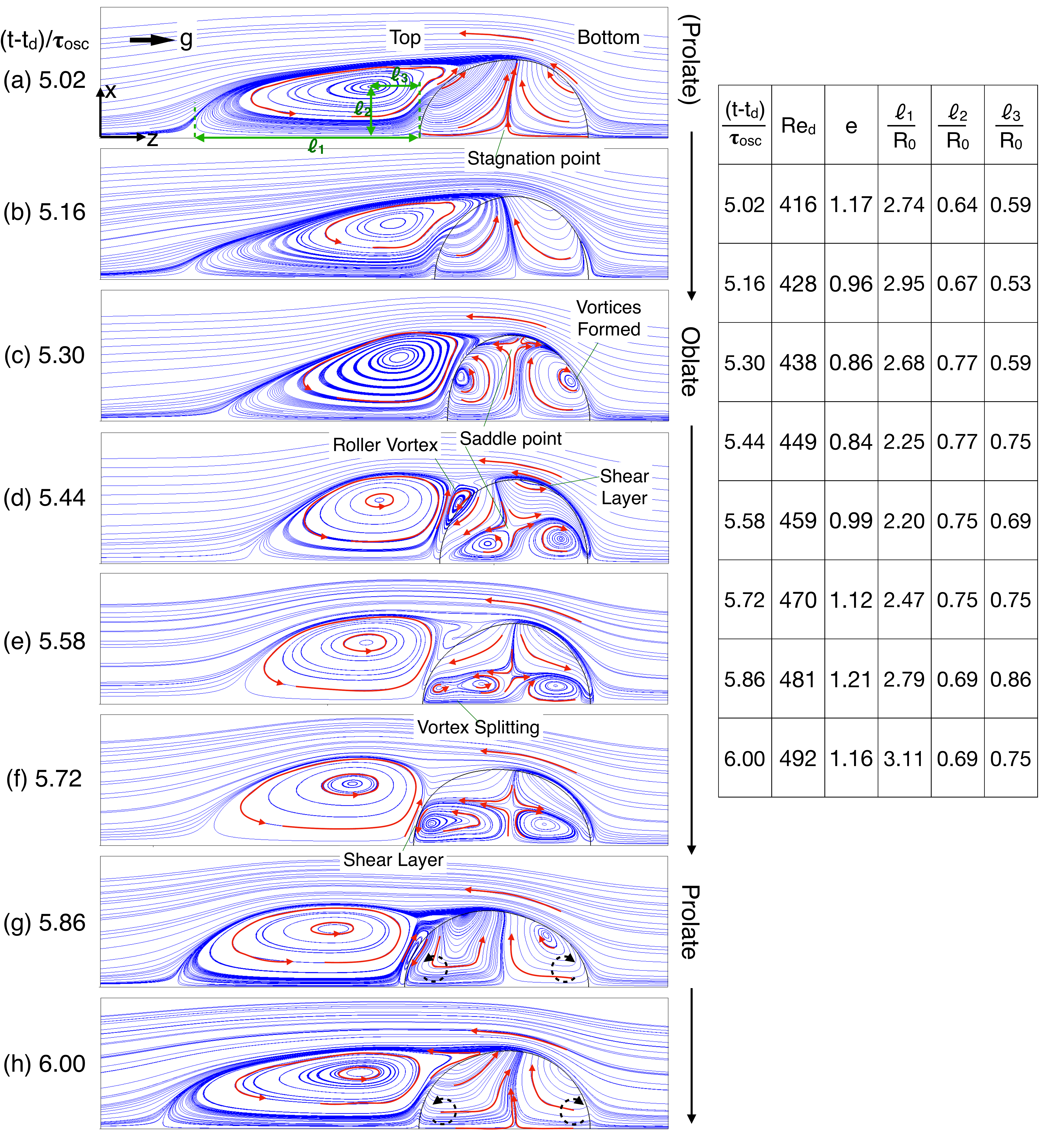}
\caption{Flow field near the oscillating and falling drop for $(t-t_d)/\tau_{osc}$ from 5 to 6. The characteristic length scales for the wake geometry, including the wake length $l_1$, the distance between the wake-vortex center and the axis $l_2$, and the distance between the wake-vortex center and the top of the drop $l_3$, are measured.   }
\label{fig:streamline}
\end{center}
\end{figure}

\tcbl{
\subsection{Flow patterns during one oscillation cycle}
}
The interplay between the falling motion and the shape oscillation 
creates a complicated transient flow which 
is different from either of the two limiting cases. 
The evolution of the flow is illustrated with streamlines in the drop reference frame in figure \ref{fig:streamline}. 
Since the second mode is dominant, the temporal variation of the flow pattern 
generally follows the cycle of the second-mode oscillation. 
The time range covered in figure  \ref{fig:streamline} is $(t-t_d)/\tau_{osc}\approx 5$ to 6, 
namely representing the sixth oscillation according to the second mode. 
The drop deforms from its prolate (elongated in $z$-direction) to oblate (flattened in $z$--direction) shapes 
in figures (a)-(c), reaching the most oblate shape 
at $(t-t_d)/\tau_{osc}\approx 5.30$. 
Then the drop returns back to the prolate shape from (c)--(g), until a new cycle starts. 

When the drop deforms from the prolate to the oblate shapes, see figures \ref{fig:streamline}(a)-(b), 
the streamlines inside the drop are quite similar to those for the second-mode free oscillation. 
A  stagnation point is formed when the fluid moves from the two poles toward the center. 
In the ground reference frame, the fluid velocity at that stagnation point is identical to the mean falling velocity of the drop. 
A close examination further shows that the stagnation point does not generally overlap with the centroid. 
In this time range, the external flow going over the drop is already strong enough to overcome the rotational flow 
induced by drop oscillations, therefore, the vortices outside the drop that are seen in the free oscillation 
(see figure \ref{fig:streamline_sketch}(a)) become invisible. At the colatitude $\theta$ about 45 and 135 degrees,
the streamlines inside the drop align well with those outside. The internal flow corresponding to the 
prolate-to-oblate oscillation is enhanced by the external flow. 

After the drop reaches the most oblate shape and starts to deform back (figures \ref{fig:streamline}(c)-(g)), 
the internal flow field becomes very different from that for the free oscillation shown in figure \ref{fig:streamline_sketch}(a). 
For the freely-oscillating drop, while the drop deforms from the oblate to the prolate shapes, 
the flow moves from the lateral side to the  the stagnation point and then bifurcates toward the two poles (see figure \ref{fig:streamline_sketch}(a)). 
However, for the falling drop, as the original internal flow due to prolate-to-oblate oscillation is 
strengthened by the external flow, the oblate-to-prolate oscillation fails to reverse flow direction 
near the stagnation point. 
Indeed, the flow direction near the stagnation point does not change through the  oscillation cycle. 
While the interface at the lateral side of the drop retracts toward the axis, the flow near the stagnation point still
tries to move toward the lateral side. As a consequence, a saddle point (a saddle curve due to the axisymmetric geometry) 
is formed, which in turn induces two vortices (vortex tubes in the axisymmetric geometry) within the drop, see figure \ref{fig:streamline}(c). 

As the drop continues to deform towards the prolate shape, the saddle point is further pushed toward the $z$--axis, 
so are the two vortices. Furthermore, as the internal circulations near the top and bottom of the drop
are not aligned with the wake and the external flow, see figure \ref{fig:streamline}(d),  
roller vortices are formed outside the drop \citep{Bergeles_2018a}. 
When the drop becomes more prolate, the two vortices inside are further flattened. 
At a certain point, see figure \ref{fig:streamline}(e), the internal vortex near the top of the drop splits into two. 

After reaching the most prolate shape, the drop starts to deform back toward the oblate shape. In this process, 
as shown in figure \ref{fig:streamline}(g), the two vortices inside the drop near the axis become invisible. 
However, they still exist, as will be shown later with vortex-identification techniques. It is just that 
the potential flow induced by the drop oscillation is so strong that, the local swirling motion cannot be shown 
by streamlines. Two new transient vortices are formed inside the drop near the lateral side, which vanish very soon. 
Then the internal flow pattern returns to the form similar to the beginning of the cycle. 

Within the time range considered  the drop oscillation is still quite strong, \eg, the second-harmonic-mode  
amplitude remains larger than 0.1 as shown in figure \ref{fig:mode}. 
As a result, the fluid inertia due to oscillation plays a significant role in the transient flow inside the drop. 
It is important to note that the internal flow pattern observed here is substantially different from Hill vortex, 
which corresponds to the long-term behavior when the drop oscillations are damped. 
In particular, the two vortices formed during the oblate-to-prolate process rotate in opposite directions compared to the 
corresponding external flows. Roller vortices are then formed in between 
the internal and external flows to satisfy the fluid kinematics. 

\tcbl{
The formation of the saddle point during the oblate-to-prolate deformation is an important feature, which is due to the different directions of the flows induced by the external shear and the shape oscillation. Therefore, the Strouhal number,  $Sr=u_{osc}/u_{ic}$, can be defined to characterize the formation of the saddle point, where $u_{osc}$ and $u_{ic}$ represent the characteristic velocities for the internal flows induced by the shape oscillation and by the external flow, respectively. While $u_{osc}$ can be estimated as $u_{osc}\approx a_2 \omega_2$, where $a_2=A_2 R_d$ and $\omega_2$ are the oscillation amplitude and frequency corresponding to the dominant second mode, $u_{ic}$ can be approximated as $u_{ic}\approx u_d \nu_{ic}$, where $\nu_{ic}$ is the internal circulation intensity \cite{Feng_2010a}. The Strouhal number can be rewritten as $Sr=a_2 \omega_2/(u_d \nu_{ic})$. When $Sr \to 0$, the droplet falls without oscillation (see Fig.\ \ref{fig:streamline_sketch}(b)). When $Sr \to \infty$ the drop oscillates without translational motion (see Fig.\ \ref{fig:streamline_sketch}(a)). For both these asymptotic limits, there is no saddle point in the flow. The saddle point will arise only when $Sr \sim O(1)$, namely when $u_{osc}$ and $u_{ic}$ are comparable.
}

\subsection{Wake topology evolution}
The characteristic length scales for the wake geometry, including the wake length $l_1$, the distance between the wake-vortex center and the axis $l_2$, and the distance between the wake-vortex center and the top of the drop $l_3$, are measured over an oscillation cycle $(t-t_d)/\tau_{osc}=5$ to 6  and are shown in figure \ref{fig:streamline}. The simulation results show that the wake length $l_1$ generally increases over time, which is consistent from former observations by \citet{Bergeles_2018a}.  At $(t-t_d)/\tau_{osc}=5$ and 6, the drop eccentricity, $e$, are the same, while the wake length increases from $l_1/R_0=2.74$ to 3.11 due to increasing $Re_d$. The values here are larger than those obtained by \citet{Bergeles_2018a} because of the larger $Re_d$. At $(t-t_d)/\tau_{osc}=5$, $Re_d=416$ and $l_1/R_0=2.74$, compared to $l_1/R_d=2.2$ for the maximum $Re_d=273$ in the former study \citep{Bergeles_2018a}.  

Furthermore, due to the higher resolution in the present simulation, variation of $l_1$ following the dominant second mode oscillation is observed, which was not shown in the former study  \citep{Bergeles_2018a}. It can be shown that $l_1$ decreases when the drop deforms from prolate to oblate shapes, and increases when the drop returns back to the prolate shape. There exits a small time lag between the temporal variation of $l_1$ and $e$ due to the inertial effect. Here $e$ reaches the local minimum at about $(t-t_d)/\tau_{osc}=5.44$ while $l_1$ does not get to the local minimum until about $(t-t_d)/\tau_{osc}=5.58$. The distance between wake-vortex center and the top of the drop $l_3$ also generally increases over time similar to $l_1$, though the increase is more gradual. As a result, its variation within the time range shown in figure \ref{fig:streamline} is mainly dictated by the drop oscillation. The amplitude increase of $l_2$ over a cycle is also small, similar to $l_3$.  The difference between $l_2$ and $l_3$ is that $l_2$ is large when the drop is oblate and is reduced when the drop turns back to the prolate shape. This is because when the drop deforms toward the oblate shape, the wake-vortex center is also pulled toward the lateral side.

\subsection{Vortex dynamics}
\begin{figure}
\begin{center}
\includegraphics [trim=0 0.1in 0 0.1in,clip, width=0.85\columnwidth]{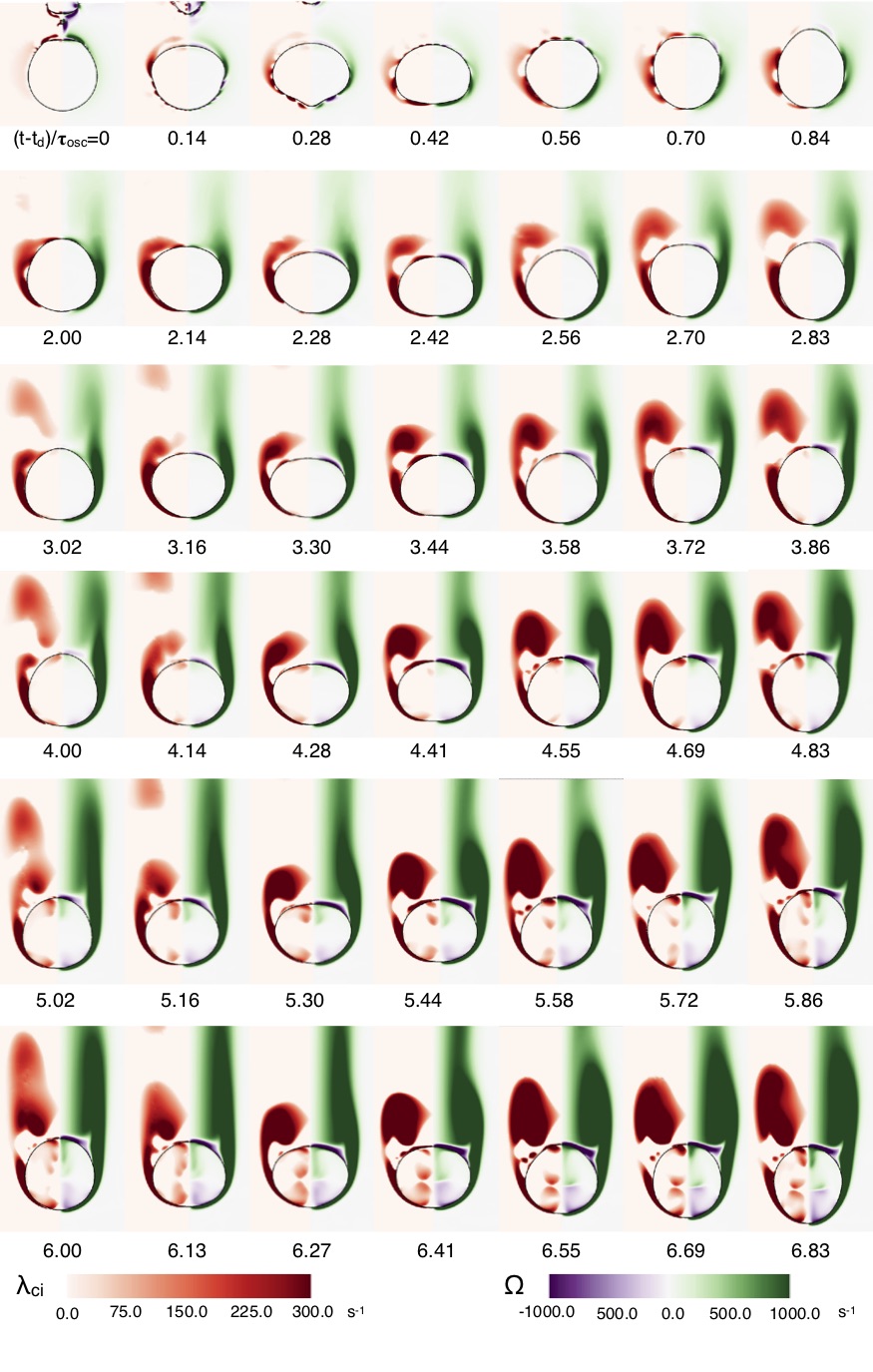}
\caption{Evolution of $\lambda_{ci}$ (left) and vorticity (right) for the dripping drop. The vortices are visualized by the $\lambda_{ci}$ criterion. }
\label{fig:lci_vort}
\end{center}
\end{figure}

%The flow around the drop is transient and is accompanied by complex vortex dynamics, 
%following the dominant second-mode oscillation cycle. 
It is well known that streamlines are insufficient to fully identify vortices. 
Galilean invariant flow properties must be used instead. 
The swirling-strength vortex-identification criterion  \citep{Zhou_1999a}, also known as
$\lambda_{ci}$-criterion, is employed here 
to illustrate the evolution of vortices, see figure \ref{fig:lci_vort}. 
The $\lambda_{ci}$-criterion has been shown to be an effective way to visualize vortices \citep{Zhou_1999a,Chakraborty_2005a}.
%\tcbl{and a brief summary of the calculation of $\lambda_{ci}$ is given in appendix \ref{sec:lambda_ci}. }
The vorticity, though which cannot fully identify the vortices as $\lambda_{ci}$ 
(since $\lambda_{ci}$ excludes the contribution from strain), is also plotted here to indicate 
the rotation directions of vortices. 
The vortex rotation directions are clockwise and counter-clockwise for $\Omega<0$ (purple color) 
and $\Omega>0$ (green color) on the right half of the drop, respectively. 

The figures are organized in such a way that the six rows represent the first, third, fourth, fifth, sixth and  seventh 
oscillations based on the dominant second mode (see figure \ref{fig:oscillation}), as reflected by the time normalized 
by the dominant second-mode period, $\tau_{osc}=\tau_2$.  

For the first row of the figure, the drop relative velocity is small and the effect of the falling motion
is negligible. The multiple small vortices outside  the drop are generated due to higher-order oscillation modes
(see also the velocity field in figure \ref{fig:multimode}). 
As time elapses, the amplitudes of the oscillations decrease in time due to viscous dissipation of the internal flow. 
It is shown in figure \ref{fig:mode} that the decay rate is faster for the higher-order modes. 
As a result, the small vortices outside the drop disappear in the second row of figure \ref{fig:lci_vort}. 
Only the larger vortices corresponding to the lower-order modes (\eg, $n\le 3$) survive. 

In the first two rows (the first three second-mode oscillations), there is no vortex 
seen inside the drop. As the falling velocity continues to increase, the influence 
of the external flow becomes stronger and vortices inside the drop start to arise,  
at about the middle of third row of figure \ref{fig:lci_vort}, ($(t-t_d)/\tau_{osc} \approx 3.5$. 
As explained above the formation of vortices inside the drop occurs when the drop deforms from 
oblate to prolate shapes and is the outcome of the interaction between drop shape oscillation and the external flow. 
The two internal vortices near the top and the bottom rotate in different directions, as indicated by the 
different colors in the vorticity plots. 

It can be seen from figure \ref{fig:Reynolds} that the drop Reynolds number reaches 190 at $(t-t_d)/\tau_{osc} \approx 2$ and the wake developing at the downstream side of the drop can be seen from the second row of figure \ref{fig:lci_vort}. 
From the subsequent rows of the figure, it can be observed that the shape and relative location 
of the wake vortex change periodically following the dominant second-mode oscillation. 

An important observation from the $\lambda_{ci}$ plots is that the vortices inside the drop 
indeed remain even when the drop shape changes from prolate to oblate, 
even though they are invisible in the streamline plots as shown in figure \ref{fig:streamline}. 
The potential flow induced by the prolate-to-oblate oscillation is strong and dominates the streamline pattern. 
Therefore, though local swirling motions exist,  they can only be shown by Galilean-invariant vortex-identification scalars 
like $\lambda_{ci}$. 
From the vorticity plots, it is learned that the rotation directions of the internal vortices do not change over 
an oscillation cycle, even though the potential flow direction changes in the second-mode oscillation cycle, 
see figure \ref{fig:streamline}. On the right half of the figure the top vortex 
always rotates in counter-clockwise direction, while the bottom one swirls in the clockwise direction all the time. 

\begin{figure}
\begin{center}
\includegraphics [width=0.9\columnwidth]{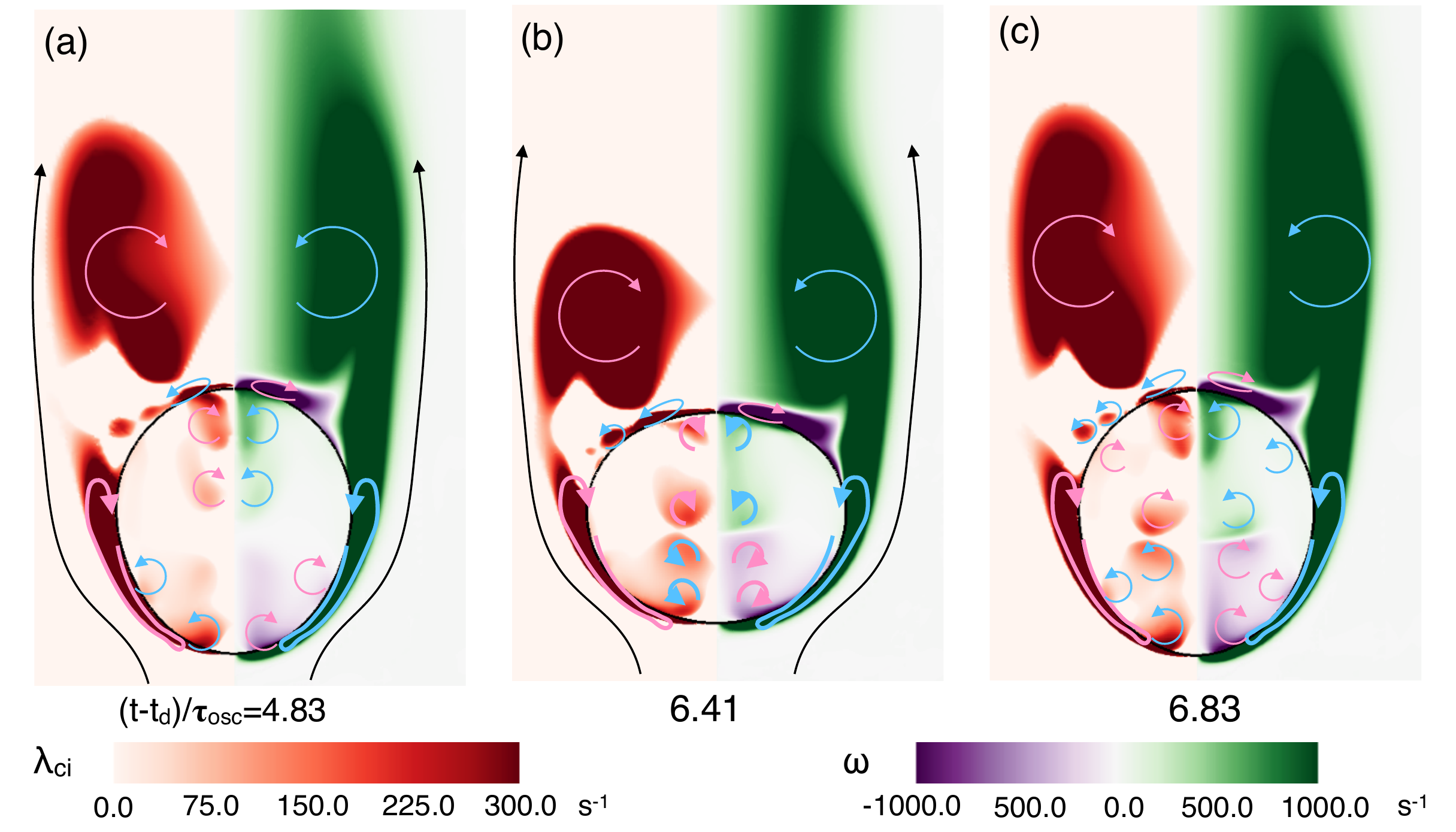}
\caption{Closeup of the vortices formed around the drop. Annotations are added to indicate the rotation direction.} 
\label{fig:lci_vort_skem}
\end{center}
\end{figure}

Closeups of the vortices with annotations are shown in figure \ref{fig:lci_vort_skem}. 
The topology of the vortices inside the drop changes within an oscillation cycle.  
When the drop deforms toward the prolate shape, the vortices are stretched and can even split into two pieces.  
During the oblate-to-prolate deformation, the vortices at the lateral side are pushed toward the axis and 
will eventually merge with the ones which are already there.  

\subsection{A summary of transient flow development inside the drop}
With the assistance of both the streamlines and contours of $\lambda_{ci}$ and vorticity, 
the development of the transient flow and the vortices interaction inside the drop can be described as follows:
\begin{enumerate}
	\item The internal flow induced by prolate-to-oblate oscillation is aligned and enhanced by the external flow. 
	\item As the falling velocity increases, at a certain point, the oblate-to-prolate deformation fails to fully 
	reverse the internal flow induced by its prolate-to-oblate counterpart. 
	\item Then a saddle point (curve) arises inside the drop when the drop deforms from its most oblate shape toward the 
	prolate shape. 
	\item The saddle point induces two vortices rotating in different directions inside the drop.
	\item As the internal circulations are different from the external flows, roller vortices are formed to satisfy 
	kinematics. 
	\item As the drop continues to deform toward its most prolate shape, the two vortices are pushed 
	toward the axis. (If there are vortices already near the axis, the new ones will merge with the old ones.)
	\item The vortices near the axis will be stretched and may split when the drop deforms toward 
	the prolate shape. 
	\item When the drop deforms back to the prolate shape, the two vortices remain present and the rotation directions 
	do not change. 
	\item Going back to (iii) and a new cycle starts. 
\end{enumerate}

%\begin{figure}
%\begin{center}
%\includegraphics [trim=0in 0in 0in 0in,clip, width=1\columnwidth]{lci_vec}
%\caption{Evolution of vortices and velocity vectors for the oscillating drop illustrating the internal circulation. The vortices are visualized by the $\lambda_{ci}$ criterion. The three rows here correspond to the fifth, sixth, and seventh oscillations according to the dominant 
%second mode shown in figure \ref{fig:oscillation}.} 
%\label{fig:lci_vec}
%\end{center}
%\end{figure}
%

\subsection{Passive scalar transport within the drop}

\begin{figure}
\begin{center}
\includegraphics [width=1.0\columnwidth]{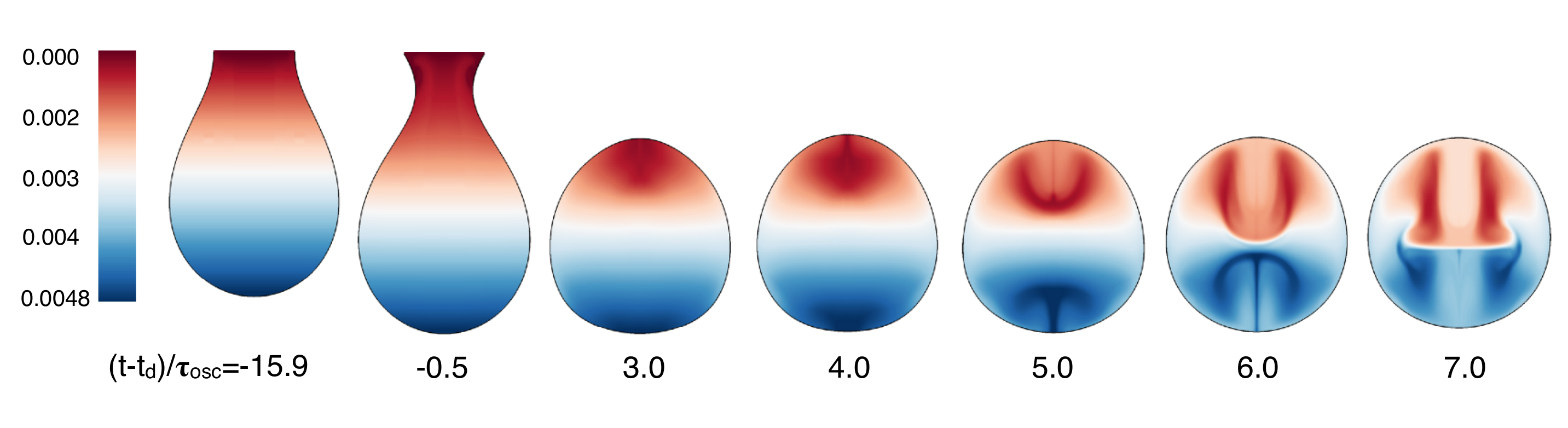}
\caption{Evolution of the tracer function distribution.} 
\label{fig:T1_iso}
\end{center}
\end{figure}

It is of interest for many drop applications to know the influence of the transient flow within a oscillating drop on scalar transport inside the drop. The question of interest is whether mixing will occur if inhomogeneous fluids are injected into the drop through the nozzle. Although mixing of different fluids inside the drop is not the focus of the present study, here a passive tracer function is introduced in the simulation to illustrate the transport process within the drop. The initial value of the tracer function is set as the streamwise coordinate $z$. 
The evolution of the tracer field serves to reveal the accumulation effect of the transient internal flow development described above 
on scalar transport. 

The advection equation of the tracer function is only solved within the liquid phase. The Godunov advection scheme with the second-order centered estimate for the velocity gradient was used. There exists a small numerical diffusion, but due to the fine mesh used, the numerical diffusion effect on the advection process is small. The results of the tracer function at different times are shown in figure \ref{fig:T1_iso}. Before the drop detaches from the nozzle, the tracer function only varies with $z$. The tracer function here can be considered to mimic an imaginary experiment in which the fluid fed in the nozzle is dyed sequentially with blue, white, and red colors. When the neck of the pendant drop develops, the tracer function is redistributed by the vortex ring created by the Ventruri jet through the neck \citep{Hoepffner_2013a}. The tracer function in the lower part of the drop remains unchanged.  

The snapshots of the drop after detachment are chosen to exhibit similar eccentricity, namely similar phases in the second-mode oscillation. When the drop simply oscillates at early time ($0<(t-t_d)/\tau_{osc} <3$), the tracer function distribution varies only in $z$, similar to the initial distribution. \tcbl{The shape oscillation by itself may introduce longitudinal motion (for example by the odd modes), 
but will not lead to net longitudinal transport of the tracer function. This is simply because the fluid motion induced by small-amplitude oscillation is symmetric and after one oscillation cycle the scalar function distribution will return to its original state.}
As the falling velocity increases, the external flow develops and interacts with the drop oscillation. Vortices arise inside the drop and they translate and interact following the drop oscillation cycle. Then stretching and folding of the fluids of different tracer function values are observed. 
As the top and bottom circulations are of different directions, the folding directions of the red and blue fluids are different. 
Though the fluids are ``mixed" inside the top and bottom portions of the drop, the two portions 
remain segregated most of the time.  
At later time, however, more complex distorted patterns of the tracer function arise, which is due to the 
unsteady motion of the saddle point (see figure \ref{fig:streamline}(c--f)). If the simulation was run for a longer time
to allow more oscillation cycles,  chaotic mixing \citep{Aref_1986a, Angilella_2003a} of inhomogeneous fluids 
may arise. More detailed investigation of transport phenomena will be left for our future work. 

 \section{Conclusions}
 \label{sec:conclusions}
The short-term transient falling dynamics of a dripping water drop has been studied. 
One specific case with a low inflow rate in the dripping regime is considered. 
The focus is on the short term behavior and the time range considered covers 
about eight dominant second-mode oscillations of the drop after it is formed. 
A high-resolution numerical simulation has been performed to investigate the oscillation and falling dynamics. 
Experiment under the same conditions was also conducted for validation purpose. 
The grid-refinement study and the excellent agreement between simulation and experiment/theory 
verify and validate the simulation results. 
Despite the low fluid inertia, the post-formation state of the drop still triggers a nonlinear oscillation. 
To rigorously account for the effect of drop formation on shape oscillation, 
the overall process including the drop growth, pinch-off, and fall, is studied. 
The interaction between the shape oscillation and the falling motion 
introduces complex oscillation dynamics and transient flow around the drop. 

\paragraph{Drop formation} The experimental results for the growing pendant drop, 
such as the relation between drop height and volume, agree well with the static pendant drop theory, 
which confirms that the drop development process is quasi-static and can be fully described by the static theory. 
This justifies the way the simulation setup by using the static pendant drop solution slightly ahead of the pinch-off time 
as the initial condition in the simulation. 
The computed drop contours for the drop growth and formation match very well with the experimental results, 
validating the setup of the numerical model. Though pinching dynamics is not the focus of the present study, 
evolutions of the velocity and pressure fields are presented to illustrate important features for low-viscosity
liquid drop formation, including the shifting of the minimum radius to the two ends of the liquid bridge, 
the interface overturning before pinch-off occurs, and the formation of the secondary drop. 
The temporal evolution of the liquid bridge minimum radius shows an initial inertial regime ($(t_d-t)^{2/3}$ power law) 
which later transitions to the viscous regime ($(t_d-t)^{1}$ linear law). The results affirm that the drop formation is precisely captured. 

\paragraph{Effect of drop formation on drop oscillation}
The post-formation state serves as the initial condition for the subsequent oscillation of the drop.  
The initial shape of the drop when it is just formed is decomposed into spherical harmonic modes. 
The initial mode amplitudes, characterized by the Fourier-Legendre coefficients, 
are found to be finite for the modes $n\le10$ considered. 
The pinching dynamics such as interface overturning introduces small-scale variation on the drop contour, 
which in turn contributes to the finite amplitudes of the higher-order modes. 
Furthermore, during the pinching process the high pressure in the neck expels fluids toward the to-be-formed drop, 
which leads to a significant downward velocity in the top region of the drop when it is just detached. 
The initial kinetic energy is as important as the initial surface energy contained in the drop shape, and is found 
to amplify the initial oscillation amplitude and to induce a phase shift in the oscillation of all the modes. 
By incorporating both the initial surface and kinetic energy, the linear model for a free drop oscillation yields 
very good predictions for the second and third modes. 

\paragraph{Effect of nonlinear dynamics on drop oscillation} 
The post-formation state of the drop triggers a moderately nonlinear drop oscillation. 
The oscillation amplitude for the dominant second mode is about 10\%, so the influence of 
finite amplitude on oscillation frequency is small for all the modes considered here. 
Nevertheless, typical nonlinear effects including asymmetry in oscillation amplitude
and  interaction between different modes are identified. The nonlinear effects are more profound 
for higher-order modes ($n\ge 4$) than lower-order modes ($n=2,3$). 
Since the majority of energy is stored in the lower-order modes, the small energy transfer between 
modes may be significant for the higher-order modes but will have little impact on the lower-order modes. 
Mode coupling is clearly reflected in the frequency spectra of the Fourier-Legendre coefficients. 
In the spectrum of a given mode $n$, a primary frequency that is very similar to the Lamb frequency
can be identified. Furthermore, the spectrum shows secondary frequencies corresponding to 
different modes due to mode coupling.  
Due to the low viscosity of water, there exists a commensurate 
relation between the $n=2$ and $4$ modes, which explains why nonlinear effects 
are always strongest for the $n=4$ mode. 

\paragraph{Effect of falling motion on drop oscillation}
The present results indicate that the effect of the fall on the oscillation frequency 
is little for the time range considered here. 
The oscillation frequency for the falling drop agrees well with Lamb's prediction 
even when the drop Reynolds number exceeds the oscillation Reynolds number for 75\%. 
This conclusion is true for both lower and higher order modes. 
The effect of the drop fall on shape oscillation lies mainly in the time evolution of the amplitudes of the various 
shape oscillation modes. 
The increasing shear stress induced by the falling motion changes the force balance with 
surface tension, resulting in a strengthened upward shift in oscillation 
amplitude for the higher-order even modes. 
The drop falling motion also seems to provide energy to the oscillations, and as a result, 
the damping in amplitude is slowed down for $(t-t_d)/\tau_{osc}\gtrsim 4$. 

\paragraph{Effect of drop oscillation on transient flow development} When the drop falls without oscillation, 
the external shear induced by the falling motion will induce the Hill vortex within the drop. 
For the present case, nonlinear shape oscillation interacts with the external flow 
induced by the falling motion, resulting in a complicated transient flow around the drop. 
When the drop oscillates from prolate to oblate shapes, the flow induced by the oscillation is aligned 
with the external flow. In contrast, for a oblate-to-prolate deformation, the flow goes against the 
external flow. As a result, a saddle point (curve for the axisymmetric geometry) arises 
in the drop, which gives rise to two counterrotating vortices. 
The rotating directions of the vortices remain unchanged, while the potential flow directions vary 
due to the dominant second-mode oscillation. The drop oscillation also influences 
the wake geometry. 
The swirling-strength vortex-identification criterion ($\lambda_{ci}$) and the vorticity are employed 
to better elucidate the vortex dynamics. When the drop oscillates, the vortices inside can be stretched
and even split. Finally, a tracer function is introduced to demonstrate the scalar transport within the drop. 
Pure shape oscillation does not induce net longitudinal transport of the tracer function. 
Stretching and folding of the scalar function contours are only observed after vortices arise within the drop. 
The unsteady motion of the saddle point creates a more distorted tracer function field, which 
may result in a chaotic mixing of inhomogeneous fluids inside the drop. 
Yet a longer simulation than the present one will be required to fully verify this.

\section*{Acknowledgement}
This work was initiated with the support of the MOST-CNRS project. The subsequent investigation was supported by the startup fund at Baylor University. YL was also supported by National Science Foundation (NSF \#1853193). The simulations were performed on the Baylor cluster \emph{Kodiak} and the simulation results are visualized by the software, \emph{VisIt}, developed by Lawrence Livermore National Laboratory. YL would also acknowledge Dr.~S.~Balachandar for helpful discussions on vortex identification 
and chaotic mixing. 

\appendix
\section{Pendant drop theory}
\label{sec:pendant_drop_theory}
 The shape of a static pendant drop can be calculated based on the equilibrium equation \citep{Padday_1973a}
\begin{equation} 
	\sigma\left(\frac{1}{\mathcal{R}_1}+\frac{1}{\mathcal{R}_2}\right) = 2{\sigma \kappa_b}-(\rho_l-\rho_g)gz'\, ,
	\label{eq:equil_drop}
\end{equation} 
where $\rho_l$ and $\rho_g$ are the water and air densities respectively and $\kappa_b$ is the curvature at the bottom of the pendant drop. The two principal radii of curvature, $\mathcal{R}_1$ and $\mathcal{R}_2$, can be calculated as 
\begin{align} 
	\frac{1}{\mathcal{R}_1} = \frac{\partial \phi}{\partial s}\,,
	\frac{1}{\mathcal{R}_2} = \frac{\sin \phi}{x'},
\end{align} 
where $s$ is the curvilinear coordinate starting from the bottom of the drop, see figure \ref{fig:static}.  
Then Eq.\ \eqr{equil_drop} can be written as 
\begin{equation} 
	\frac{\partial \phi}{\partial s} =2 \kappa_b-\frac{(\rho_l-\rho_g)gz'}{\sigma} - \frac{\sin \phi}{x'}. 
	\label{eq:equil_drop_2}
\end{equation}  

It can also be shown from geometry that
\begin{align} 
	\frac{\partial x'}{\partial s} & = \cos \phi\, ,\label{eq:geo_1}\\
	\frac{\partial z'}{\partial s} & = \sin \phi\, . \label{eq:geo_2} 
\end{align} 
At the end the ODE system, Eqs.\ \eqr{equil_drop_2}, \eqr{geo_1}, and \eqr{geo_2}, can be solved numerically to yield the contour of the static pendant drop. 

\section{Grid independence study for the evolution of the amplitude of spherical harmonic modes }
\label{sec:mode_grid}
\begin{figure}
\begin{center}
\includegraphics [width=0.99\columnwidth]{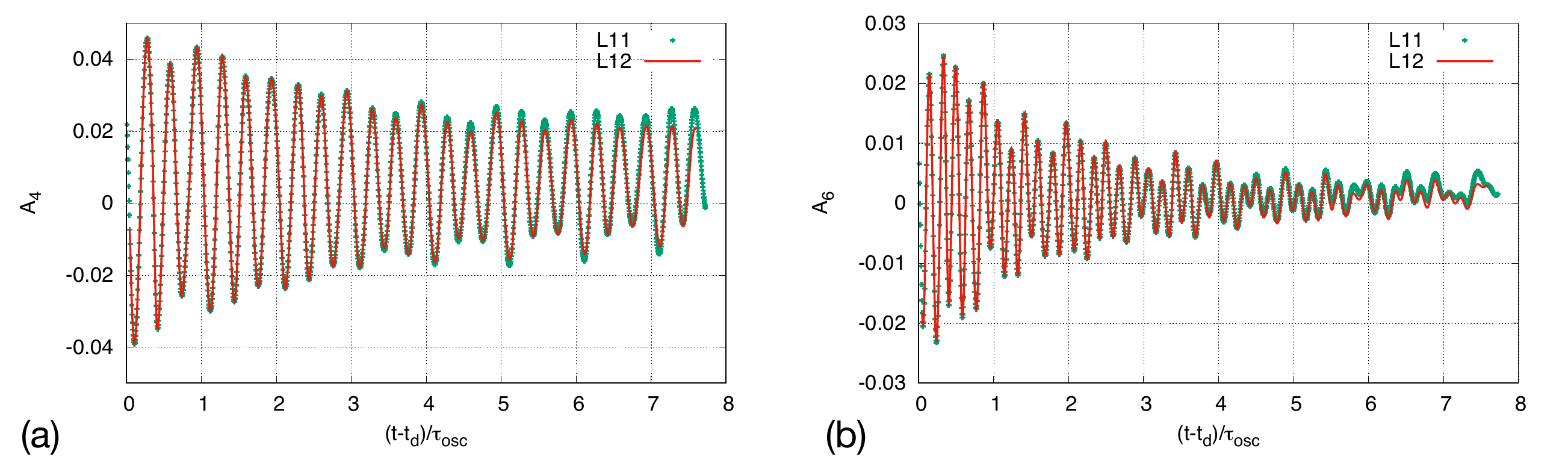}
\caption{Temporal evolutions of the Fourier-Legendre coefficients for the (a) $n=4$ and (b) $n=6$ modes for different mesh resolutions.} 
\label{fig:mode_grid}
\end{center}
\end{figure}

To fully confirm the results of Fourier-Legendre coefficients shown in figure \ref{fig:mode} are grid independent, 
a simulation with an additional refinement level $L=12$ has been performed. The results for the $n=4$ and 6 modes 
are shown in figure \ref{fig:mode_grid}, confirming that the important conclusions made related to the effect of drop formation, 
nonlinear dynamics, and falling motion on the drop oscillation are independent of the grid resolution. 

%\section{Vortex identification through $\lambda_{ci}$-criterion}
%\label{sec:lambda_ci}
%The swirling-strength vortex-identification criterion ($\lambda_{ci}$-criterion) \citep{Zhou_1999a} is used in the present study to illustrate the evolution of vortices within the drop when it falls. The $\lambda_{ci}$-criterion is calculated based on the velocity gradient tensor $\nabla \bs{u}$. The real and complex conjugate eigenpairs of $\nabla \bs{u}$ are $(\lambda_r, \bs{v}_r$) and $(\lambda_{cr} \pm \mathrm{i} \lambda_{ci}, \bs{v}_{cr}\pm \bs{v}_{ci})$. 
%The flow in the plane spanned by $(\bs{v}_{cr} , \bs{v}_{ci})$ is locally swirling. The time for one swirling cycle along the stream line is $2\pi/\lambda_{ci}$, therefore, $\lambda_{ci}$, referred to as ``swirling strength", serves as a measure of the local swirling rate inside the vortex \citep{Chakraborty_2005a}. In the present problem, the two-dimensional $\lambda_{ci}$-criterion is used and the expression was given by \cite{Chen_2015d}: 
%\begin{align}
%	\lambda_{ci} = \frac{1}{2}\sqrt{-4\left( \pd{u}{z} \pd{w}{x}  -  \pd{u}{x} \pd{w}{z}\right) - \left(\pd{u}{x}+\pd{w}{z}\right)^2}.
%\end{align}


\end{document}